\newif\ifSC
\SCtrue
\ifSC
\documentclass[onecolumn,draftclsnofoot,12pt]{IEEEtran}
\else
\documentclass[twocolumn,10pt]{IEEEtran}
\fi

\usepackage[utf8]{inputenc}
\usepackage{etoolbox}

\newtoggle{SC}
\togglefalse{SC}
\ifSC
\toggletrue{SC}
\fi

\usepackage{amsmath,amssymb,amsthm,mathrsfs,amsfonts,dsfont}
\usepackage{verbatim}
\usepackage{graphicx}
\usepackage{color}
\usepackage{epstopdf}
\usepackage{url}

\usepackage{tabulary,bbm}
\usepackage{cite}

\newcommand{\prob}[1]{ \mathbb{P}\left[#1\right] }
\newcommand{\expect}[1]{\mathbb{E}\left[#1\right] }
\newcommand{\expects}[2]{\mathbb{E}_{#1}\left[#2\right] }
\newcommand{\laplace}[1]{\mathcal{L}_{#1} }

\newcommand{\losind}[0]{ {s_{ki}} }
\newcommand{\Ball}[0]{ \mathcal{B} }
\newcommand{\dB}{\text{dB}}
\newcommand{\SINR}{\text{SINR}}

\newcommand{\equalspace}{\ \ \ \ }
\newcommand{\ie}{{\em i.e. } }
\newcommand{\sco}{s'}

\iftoggle{SC}{
\newcommand{\includefig}[1]{\includegraphics[width=0.57\textwidth]{figures/#1}\vspace{-.15in}}}
{\newcommand{\includefig}[1]{\includegraphics[width=0.95\columnwidth]{figures/#1}\vspace{-.15in}}
}

\newcommand{\includefigtrimmed}[2]{\includegraphics[width=0.88\textwidth,trim=#2,clip=true]{figures/#1}}

    \newcommand{\tho}{$^\text{th}$ }
    \newcommand{\UE}{$\mathrm{UE}_0$}
     \newcommand{\BS}{\mathrm{BS} }

\newcommand{\x}{\mathbf{x}}
    
     \newcommand{\figurekeyword}{Fig. }
        \newcommand{\Noise}{\sigma^2}
    \newcommand{\thetab}{\theta_\mathrm{b}}
\newcommand{\K}{M}
\renewcommand{\L}{\mathrm{L}}
\newcommand{\N}{\mathrm{N}}
\newcommand{\Pc}{\mathrm{P}}
\renewcommand{\a}{\alpha}
\newcommand{\Rc}{\mathrm{R}}
\newcommand{\intd}{\mathrm{d}}
\newcommand{\rthres}{\rho}

\newcommand{\user}{\mathrm{u}}

\newcommand{\Pcu}[1]{\mathrm{P}^\mathrm{c}_{#1}}

\newcommand{\VoidF}[2]{\mu_{#1}{\left(#2\right)}}

\newtheorem{lemma}{Lemma}
\newtheorem{theorem}{Theorem}
\theoremstyle{definition}

\newtheorem{corollary}{Corollary}[theorem]

\newcommand{\addvzerothree}[1]{#1}
\newcommand{\addvzerofour}[1]{#1}
\newcommand{\addvzerosix}[1]{#1}

\newcommand{\addvone}[1]{#1}

\title{On the Feasibility of Sharing Spectrum Licenses  in mmWave Cellular Systems}

\author{Abhishek K. Gupta, Jeffrey G. Andrews, Robert W. Heath, Jr.
\thanks{A. K. Gupta ({\tt g.kr.abhishek@utexas.edu}), J. G. Andrews ({\tt  jandrews@ece.utexas.edu}) and  R. W. Heath Jr.  ({\tt rheath@utexas.edu}) are with Wireless Networking and Communications Group, The University of Texas at Austin, Austin, TX 78712 USA. This work was supported by the National Science Foundation  under Grant 1514275.}
\thanks{\addvone{A part of this paper is presented at  ITA Workshop, Feb. 2016 in San Diego, CA, USA\cite{GuptaAndHeath2016}}.}
}

\begin{document}
\maketitle
\begin{abstract}
The highly directional and adaptive antennas used in mmWave communication open up the possibility of uncoordinated sharing of spectrum licenses between commercial cellular \addvone{operators}. There are several advantages to sharing including a reduction in license costs and an increase in spectrum utilization. In this paper, we establish the theoretical feasibility of spectrum license sharing among mmWave cellular \addvone{operators}. We consider a heterogeneous multi-operator system containing multiple independent cellular networks, \addvone{each owned by an operator}. We then compute the SINR and rate distribution for downlink mobile users of each network. Using the analysis, we compare systems with fully shared licenses and exclusive licenses for different access rules and explore the trade-offs between system performance and spectrum cost. We show that sharing spectrum licenses increases the per-user rate when antennas have narrow beams and is also favored when there is a low density of users. We also consider a multi-operator system where BSs of all the networks are co-located to show that the simultaneous sharing of spectrum and infrastructure is also feasible. We show that all networks can share licenses with less bandwidth and still achieve the same per-user median rate as if they each had an exclusive license to spectrum with more bandwidth.

\end{abstract}


\def \fodname {FullContent}
\def \podname {PaperContent}

\bstctlcite{IEEEexample:BSTcontrol}

\section{Introduction}
Due to scarcity of spectrum at conventional cellular frequencies (CCF), the use of higher frequencies such as mmWave has been proposed  for 5G cellular networks \cite{mmwaveintro, Andrews5G, BaiAhmedMag14}. The mmWave cellular systems are expected to consist of multiple networks, each deployed by an independent cellular operator with possibly closed access only to its customers.  It is worth noting that communication at mmWave frequencies has non-trivial differences when compared to communication at CCF. For example, the typical use of many antennas in mmWave systems results in highly directional communication \cite{Roh2014,Rappaport2013}, and under some circumstances, it is  noise limited \cite{Akdeniz2014,Larew2013,Ghosh2014,Ab2013}. In general,  mmWave communication causes less interference to neighboring BSs operating in the same frequency bands compared to communication at frequencies below 6 GHz \cite{mmwaveintro,Bai2014}. This leads to possibility of a new kind of sharing \addvone{which we call as {\em spectrum license sharing}: independent cellular operators who own licenses for separate frequency bands agree to share the complete rights of operation in each other's bands without  any explicit coordination.} Sharing  licenses (if possible), will allow all networks to use the full spectrum simultaneously without impacting the individual achieved rates and help networks to reduce their expenses by sharing the license costs.
 Such uncoordinated sharing of licenses was not possible in conventional cellular frequencies due to high interference caused by serving BSs which renders the channel unusable to any nearby BS of other networks operating in the same frequency band. 
 

\subsection{Background and Related Work}

 \addvone{
In  conventional cellular licensing, commercial operators buy exclusive licenses where they have exclusive and complete control over a band of spectrum.  For mmWave bands, there are no regulatory rules for cellular services yet, although there are many incumbent services  including fixed services, satellite to earth communications, military, and research activities and unlicensed operations.} It has been reported that the spectrum remains underutilized \cite{FCC2002} due to this exclusive licensing at CCF which is expected to  be even worse for mmWave bands.  There has been significant work related to cognitive radios to help fill the gaps in the underutilized spectrum by letting secondary users make use of the spectrum band via  sensing based access control  \cite{Haykin2005,Kang2009,Stevenson2009,Akyildiz06,Stotas2011} for CCF. In \cite{Kamal2009, Lin2014,Karunakaran2014,Alsohaily2013}, various aspects and performance of spectrum sharing were studied in a cellular setting.  Most related to this paper, in \cite{Karunakaran2014}, dynamic spectrum sharing between different operators was shown to achieve reasonable sensing performance  in 3GPP LTE-A systems with carrier aggregation.   Various cognitive license sharing schemes such as licensed shared access (LSA) and authorized shared access (ASA) were proposed \cite{Dahlman2013,ECC205} which allow more than one entity to use the spectrum. In the presence of incumbent services,  the above mentioned techniques \cite{Dahlman2013,ECC205}  would authorize a cellular system to transmit, only when the incumbent services are idle \cite{Mueck2014,Andrews5G166, Andrews5G167}. Implementation would  require some kind of sensing or central coordination, which may waste important resources resulting in underutilization of spectrum.  Further, it was shown in \cite{Elsawy2013,Elsawy2013Mag} that spectrum gaps will be rare for ultra-dense deployment of small cells in a multi-tier network and cognitive sensing may not give the desired gain. Another way to resolve the transmission conflicts between multiple licensees (or entities in case of unlicensed spectrum) is by the use of a central database which keeps track of transmission of each licensee \cite{FCCNOI}. This reduces the requirement of  continuous listening/sensing of the spectrum by each licensee/entity but creates significant feedback/transmission overhead and possibly delay due to the central database.

 In recent work, stochastic geometry has emerged as a tractable approach to model and analyze various  wireless systems. For example, the performance of a single operator  mmWave system had been investigated in prior work using tools from stochastic geometry \cite{Akoum2012,GCBai2013, Bai2014, SinghBackHaul2015}. 
  In \cite{Akoum2012}, a stochastic geometry framework for analysis of mmWave network was proposed. In \cite{GCBai2013,Bai2014}, a blocking model for mmWave communication was proposed to distinguish between line-of-sight (LOS) or non-LOS (NLOS) transmission links and performance metrics such as coverage probability and per-user rate were derived using stochastic geometry. In \cite{SinghBackHaul2015}, a stochastic geometry framework was presented for mmWave  network with backhaul and co-existent microwave network. In \cite{Kulkarni2014}, it was shown that even a very dense mmWave network tends to be noise limited for certain choices of parameters. A major limitation of the prior work on mmWave systems  \cite{Akoum2012,GCBai2013, Bai2014, SinghBackHaul2015,Kulkarni2014} is the assumption of a single operator.  
To study the impact of  an active BS of an operator on neighboring ones of another operators, a more general framework containing multiple operators is needed. For CCF, a heterogeneous system (HetNet) consisting of multiple tiers of the same operator was studied  in \cite{DhillonHetNet} where all the tiers were operating in the same frequency band. Similarly for mmWave, a numerical framework was presented  for  a mmWave HetNet with generalized fading in \cite{Renzo2015}. In \cite{Renzo2015}, only single operator was considered and BSs were assumed to  have open access to all the users, while as envisioned in this paper, commercial providers are expected to have closed access to its customers or users.  Cognitive  networks have also been studied using tools from stochastic geometry for CCF. For example, in \cite{LeeAndrews2011}, a stochastic geometry model was presented for spectrum sharing  to characterize the system performance in terms of transmit capacity.  In \cite{Nguyen2010}, a cognitive carrier sensing protocol was proposed and analyzed for a network consisting of multiple primary and secondary users and spectrum access probabilities and transmission capacity were computed. 
\addvone{There is  limited work to analyze the impact of spectrum sharing among operators at CCF using stochastic geometry. For example, in \cite{DaSilva2015}, the impact of infrastructure and CCF spectrum sharing was studied using stochastic geometry. 
As discussed above,  communication at mmWave frequencies has different characteristics and modeling considerations when compared to CCF. Therefore, these results need to be reevaluated for mmWave systems and a new mathematical model is required 
for possible license sharing in mmWave systems.
 }
In the above mentioned prior work \cite{DaSilva2015, Akoum2012,GCBai2013, Bai2014, SinghBackHaul2015, Kulkarni2014,DhillonHetNet, Renzo2015, LeeAndrews2011, Nguyen2010}, the impact of inter-operator interference in mmWave systems and the feasibility of license sharing  are not studied, which is the main focus here.

\subsection{Contributions}
We establish the feasibility of  uncoordinated sharing of spectrum licenses in a multi-operator mmWave system. We model a multi-operator mmWave system where every operator owns a spectrum license of fixed bandwidth with a provision to share the complete rights over its licensed spectrum with other networks. Next, we compute the  performance of such system in terms of signal-to-interference-and-noise (SINR) and rate coverage probability. A main conclusion of our work is that it is  feasible to share spectrum licenses among multiple mmWave operators without requiring any coordination among them. The main contributions of this paper can be categorized broadly as follows.

\addvone{
\textbf{Modeling a multi-operator cellular system: } We model a multi-operator mmWave system using a two-level architecture,  where the system consists of multiple heterogeneous networks, each owned by an independent mmWave cellular operator. Each operator owns a license for a frequency band and its network consists of  its own BSs and users which are independent of other networks.   
This model  provides a general framework for access and license sharing by introducing the notion of access and sharing groups. The operators can form multiple access groups where users of each member of an access group can connect to  BSs of any member of the same access group. Similarly,  the operators can form sharing groups where each member of a sharing group can use the spectrum of all members of the same sharing group.} 

\textbf{Establishing the feasibility of uncoordinated sharing of spectrum licenses:} We derive the combined mean user-load on BSs of each network. We derive expressions for  SINR and rate coverage of a typical user of an operator using the tools from stochastic geometry.   Then, we compare systems with  shared licenses and  exclusive licenses and show that spectrum license sharing  achieves higher performance in terms of per user rate. We  investigate the effect of antenna beamwidth on the feasibility of license sharing and show that spectrum license sharing is more favorable as communication becomes more directional. Finally, we quantify the benefit of license sharing in term of license cost reduction and show that networks can save  significant amount of money if they  share their licenses.
 
 \addvone{
\textbf{Modeling co-located operators:} To include the impact of correlation among deployments of operators, we also consider a co-located deployment where site locations  are defined by a single PPP and each operator has one BS at each site.  We show that spectrum sharing is feasible even in this case. This  important result indicates that multiple operators can also share  site infrastructure (such as macrocell towers, as is common practice today) while  also sharing their spectrum licenses. A practical deployment with some co-location and some unique sites will lie between these two extremes of independent and totally co-located BS locations.
}
 
 The rest of the paper is organized as follows. Section \ref{Sec:SysMod} explains the system model and enumerates few special combinations of accesses and license sharing schemes. In Section \ref{Sec:SINRRate}, the expressions for SINR and rate coverage probability are derived. In Section \ref{Sec:4}, we compare the  aforementioned cases and Section \ref{Sec:PL} discusses the impact of partial loading where some BSs are turned off due to lack of  users associated to them. Section \ref{Sec:Results} presents numerical results and derives main insights of the paper. We finally conclude in Section \ref{Sec:Conclusions}.
 

\section{System Model}\label{Sec:SysMod}
We consider a system consisting of $\K$ different cellular operators  which coexist in a particular mmWave band. Each operators owns a network $\Phi_m$ consisting of its own BSs  and users. The locations of BSs are modeled using a  Poisson Point Process (PPP) with intensity $\lambda_m$ and the locations of users are modeled as independent PPP with intensity $\lambda^\user_m$. \addvone{The PPP assumption can be justified by the fact that nearly any BS distribution in 2D results in a small fixed SINR shift relative to the PPP \cite{Guo2015, GantiArxiv}. The locations of BSs of multiple operators have been modeled by superposition of independent PPPs  in the past work \cite{DaSilva2015}. Different point processes such as the Log-Gaussian Cox process can also be used to model the locations of a multi-operator system as discussed in \cite{DaSilva2015J}. For tractability we have assumed  independence among operators; but, to address the impact of correlation of locations among the operators, we also consider another case with full spatial correlation among operators where all operators are co-located with their locations defined by a single PPP. One can think of real multi-operator deployments as lying between these two corner cases of complete independence and co-location.}  The BSs of each network can transmit with power $P_m$. We denote the total spectrum by $B$ and suppose that the $m^{th}$ operator  owns  a license for an orthogonal spectrum of $B_m$ bandwidth which it can share with others. 
We  consider a typical user \UE~at the origin without loss of generality. Let us assume this is a user of the $n^{th}$ operator. 

\subsection{Channel Model}
Let us consider a link between this typical user and a BS  of network $m$ located at  $x$ distance from the origin. This link can be LOS or NLOS  link which we denote by the variable link type $s$, which can take values $s=\L$ (for LOS) or $s=\N$ (for NLOS). We assume that the probability of a link being LOS is dependent on $x$ and independent of types of other links and is given by $p(x)=\exp(-\beta x)$  \cite{Bai2014,BaiVaze2014}. 
The analysis can be  extended to other blocking models.
The path loss from the BS to the user is modeled as 
$\ell_s(x)=C_s\left(x\right)^{-\a_s}$
where $\a_s$ is the pathloss exponent and $C_s$ is the gain for $s$ type links.
\addvone{
Conditioned on this typical user, the BSs of the each network $m$ can be categorized into LOS or NLOS  based on the type of their link to this typical user. Therefore, the BS PPP of network $m$ can be seen as superposition of the following two independent (non-homogeneous) BS PPPs because of the independent thinning theorem \cite{Stoyanbook},
\begin{itemize}
\item[1.]  $\Phi_{m,\L}$ with density $\lambda_{m,\L}(x)=\lambda_m p(x)$ containing all the BSs with LOS link to  \UE,  and 
\item[2.] $\Phi_{m,\N}$ with density $\lambda_{m,\N}(x)=\lambda_m (1-p(x))$ containing all the BSs with NLOS link to \UE.
\end{itemize}}
 Note that this
results in total $2\K$ classes (known as tiers) of BSs where each tier is denoted by $\{m,s\}$. Here $m$ and $s$ represent the index of the network and the link type, respectively. The average number of BSs of tier $\{m,s\}$ in area $A$ is 
given by
$
\Lambda_{m,s}(A)= \int_A \lambda_{m,s}(\|\x\|) \intd \x$.
Therefore, the average number of LOS and NLOS BSs in a ball $\Ball(r)$ with radius $r$ for network $m$ are given by
\iftoggle{SC}{
\begin{align*}
\Lambda_{m,\L}(\Ball(r))&= 2\pi \int_0^r \lambda_{m} p(x) x\intd x=2\pi \int_0^r \lambda_{m} e^{-\beta x} x\intd x
\iftoggle{SC}{}{\\&}
=\pi\lambda_{m}\frac{2 }{\beta^2}\gamma(2,\beta r)\\
\Lambda_{m,\N}(\Ball(r))
&=2\pi \int_0^r \lambda_{m} (1-p(x)) x\intd x
\iftoggle{SC}{}{\\&}=\pi \lambda_m \left(r^2-\frac{2}{\beta^2}\gamma(2,\beta r)\right).
\end{align*}}
{
\begin{align*}
\Lambda_{m,\L}(\Ball(r))&= 2\pi \int_0^r \lambda_{m} p(x) x\intd x=2\pi \int_0^r \lambda_{m} e^{-\beta x} x\intd x
\iftoggle{SC}{}{\\&}
=\pi\lambda_{m}\frac{2 }{\beta^2}\gamma(2,\beta r)\\
\Lambda_{m,\N}(\Ball(r))
&=2\pi \int_0^r \lambda_{m} (1-p(x)) x\intd x
\end{align*}
\begin{align*}
=\pi \lambda_m \left(r^2-\frac{2}{\beta^2}\gamma(2,\beta r)\right).\equalspace\equalspace
\end{align*}
}
Let us denote the $j^{th}$ BS of network $m$ as $\BS_{mj}$.   Hence, the effective channel between   $\BS_{mj}$ and the user \UE is given as
$h_{mj} P_m \ell_{s_{mj}}(x_{mj})$ 
where $s_{mj}$ denotes the link type between the user and   $\BS_{mj}$  
and  $h_{mj}$ is an exponential random variable denoting Rayleigh fading. 
We observe and show in the numerical section that considering a more general  fading model such as Nakagami does not provide any additional design insights, but it does complicate the analysis significantly. Therefore we will consider only Rayleigh fading for our analysis. \addvzerothree{We do not consider shadowing separately as it is mostly covered by the blocking model. We  show in the numerical section that including shadowing does not change \addvzerosix{any of the observed trends.}}

\begin{table}[t!]\label{notation}
\caption{Summary of Notation}
\begin{tabulary}{\columnwidth}{ |l | L | }\hline
{\bf Notation} &{\bf Description}\\ \hline
$\Phi_m, \lambda_m$ & PPP consisting of the locations of BSs of operator $m$, BS density of operator $m$ \\ \hline
$P_m, B_m,W_m$ & Transmit power of BSs of operator $m$, licensed bandwidth of operator $m$, bandwidth available to operator $m$ via sharing of licenses \\ \hline
$\Phi^\user_m$, $\lambda^\user_m$ &  PPP modeling locations of users of operator $m$, the density of this PPP\\ \hline
$\L,\N$ &   Possible values of link type: L denotes LOS, N denotes NLOS\\ \hline
$\{m,p\}$ & Notation representing the tier having all BSs of link type $p$ of operator $m$\\ \hline
$\Phi_{mp}, \lambda_{mp}$ &  The PPP modeling the locations of BSs of the tier $\{m,p\}$, the density of this PPP\\ \hline
$G_1,G_2,\thetab$ &  BS antenna parameters: maximum gain, minimum gain and half beamwidth\\ \hline
$C_p$, $\alpha_p$  & Path-loss model parameters: path-loss gain and path-loss exponent of any link of type $p\in\{\L,\N\}$ \\ \hline
$p(r),\beta$& $p(r)$ is the probability of being LOS for a link of distance $r$,  $\beta$ is the blocking  parameter.\\ \hline
\UE & A typical user at origin for which analysis is performed.\\ \hline
$n$ & The operator which \UE belongs to\\ \hline
$S_n$ & The set consisting of all operators which a user of operator $n$ has access to\\ \hline
$\BS_{mj}, \x_{mj},x_{mj}$ & $j^{th}$ BS of the network $m$, its location (here $\x_{mj}\in \Phi_m$), its distance from the origin\\ \hline
$s_{mj}$ & Fading faced by the link between  $\BS_{mj}$ and the user, the type of this link\\ \hline
$k, i$ & The operator associated with \UE, the index of the serving BS of this operator\\ \hline
$s, x $ &Type of the link between the serving BS and the user, its distance from the origin\\ \hline
$Q_k$ &  The sharing group of operator $k$ which is also the set of all operators interfering to the operator $k$.\\ \hline
$m$ & Indices representing an operator or a network, in particular a member of set $Q_k$\\ \hline
$D^{ks}_{mp}(x)$ & Exclusion radius for BSs of tier $\{m,p\}$  when \UE is associated with BS of tier $\{k,s\}$ located at $x$ \\ \hline
$A_k^n$ &  Association probability of a user of network $n$ to be associated with BS of operator $k$ \\ \hline
$\Pcu{ks}$ &  Probability of SINR coverage of \UE when associated with tier $\{k,s\}$ \\ \hline
$N_m^u$ &  The mean number of users associated to  a BS of network $m$\\ \hline
$\kappa_m(z)$ &Probability of a BS of network $m$ having  $z$ number of associated users\\ \hline
$R_{ki}, \Rc^{\mathrm{c}}_k$ &  Instantaneous rate and rate coverage of \UE when associated with network $k$ \\ \hline
$\Rc^{\mathrm{c}}(\rthres)$ &  Rate coverage of \UE, \ie $\prob{R_k\ge \rthres}$ \\ \hline
\end{tabulary}
\end{table}

We assume that BSs of every network are equipped with a steerable antenna having radiation pattern given as \cite{Bai2014}
\begin{align*}
G(\theta)=\begin{cases}
G_1 & |\theta|<\thetab\\
G_2 & \text{otherwise }
\end{cases}.
\end{align*}
Here $G_1\gg G_2$ and $\thetab$ denotes half beamwidth. The angle between the  BS $\BS_{mj}$  antenna and direction pointing to the user \UE is denoted by $\theta_{mj}$. \addvzerothree{We have assumed the same antenna pattern for all networks \addvzerofour{to avoid complicating the expressions unnecessarily}.  
The analysis can  be extended to a system where each network has a different transmit antenna pattern.} We assume that the user is equipped with a single omni-directional antenna. {Although users will also have directional antennas, it will  be analytically equivalent to aggregating  the transmitter and receiver gains at BS antennas. Considering the UE antenna gain and the BS antenna gain separately does not change the observed trends and hence, is left for  future work. Note that we consider single stream operation in this work. More advanced mmWave cellular systems may employ massive MIMO \cite{BaiMassiveMIMO15, MazertaMassiveMIMO} or multi-stream MIMO using hybrid beamforming \cite{BaiAhmedMag14}. Generalizing to these other architectures is a topic of future work.} 

\subsection{Access and License Sharing Model}
We assume that a user of operator $n$ can be associated with any BS from a particular set of operators denoted by access set $S_n$. Two special cases of access  are open and closed. In an open access system, a user can connect to any operator and therefore $S_n=\{1,2,\cdots \K\}$. In a closed access system, a user can connect only to the operator it belongs to, and therefore $S_n=\{n\}$.

 We assume that license sharing is performed by forming mutually exclusive groups, known as {\em sharing groups}. All the operators in each group share the whole spectrum license such that each operator within a group has equal bandwidth available to it. The effective bandwidth available to each operator after sharing is denoted by $W_m$. For example, in a system of 5 operators, suppose that operators 1, 2 and 3 form a group and operators 4 and 5 are in second group. Hence, after license sharing, 1, 2 and 3 will have access to the the aggregate band of total $B_1+B_2+B_3$ bandwidth, {\em i.e. } $W_1=W_2=W_3=B_1+B_2+B_3$. Similarly operators 4 and 5  will have access to the aggregate band of bandwidth $W_4=W_5=B_4+B_5$. We denote the sharing group containing the $k^{th}$ operator by $Q_k$.
The  user \UE~experiences interference from all networks operating in the spectrum of associated operator $k$. The set of the interfering BS is equal to the sharing group containing the $k^{th}$ operator which is $Q_k$. \addvone{Note that the aggregate spectrum of a sharing group can be fully used by all members. Hence, for a particular network, the set of interfering networks remains the same for its complete available spectrum band. One  example of license sharing is  a system with fully shared licenses, in which there is only one sharing group containing all the operators and all of them can use the whole frequency band. Here, the available bandwidth $W_k$ to each operator  is  $B$ and the set of interfering networks  is $Q_k=\{1,2,\cdots \K\}$. For the case when all  operators have  exclusive licenses, there are $\K$ sharing groups each containing only one operator which indicates that no operator shares its license. Hence, the  bandwidth available to the operator $k$ is $W_k=B_k$ and the set of  networks interfering to it is $Q_k=\{k\}$}.

Now, the effective received power from a BS $\BS_{mj}$ at user \UE~is given as
\begin{align}
P_{mj}&=P_j h_{mj}\ell_{s_{mj}}(x_{mj})G(\theta_{mj}).
\end{align}
Hence, the average received power from  $\BS_{mj}$  at \UE~without the antenna gain is given by 
\begin{align}
P^\mathrm{avg}_{mj}=P_m\ell_{s_{mj}}(x_{mj}).
\label{eq:avgrxpower}
\end{align}
 We assume the maximum average received power based association in which any user associates with the BS providing highest  $P^\mathrm{avg}_{mj}$ among all the networks it has access to ({\em i.e.} access set).  Let us denote the operator the user \UE~associates with by $k$ and the index of the serving BS by $i$.
Since the serving BS aligns its antenna with the user, the angle $\theta_{ki}$ between the serving BS antenna and user direction  is $0^\mathrm{o}$ and the effective received power of this BS is given as
$P_{ki}=P_k h_{ki}\ell_{s_{ki}}(x_{ki})G(0)=P_k h_{ki}\ell_{s_{ki}}(x_{ki})G_1.$
For each interfering BS $\BS_{mj}$ where $m\in Q_k$ and $(m,j)\ne(k,i)$, the angle $\theta_{mj}$ is assumed to be uniformly distributed between $-\pi$ and $\pi$. 

Now, the SINR at the typical user \UE~ that is associated with the $i^{th}$ BS of the operator $k$ is given as
\begin{align}
\SINR_{ki}=\frac{P_k h_{ki}\ell_{s_{ki}}(x_{ki})G_1}{\Noise_k+I}
\end{align}
where $I$ is the interference from all BSs of networks in set $Q_k$ and is given by
\begin{align}
I&=\sum\limits_{m\in Q_k}\sum\limits_{j\in \Phi_{m}}P_m h_{mj}\ell_{s_{mj}}(x_{mj})G(\theta_{mj})
\iftoggle{SC}{}{\nonumber\\&}
=\sum\limits_{m\in Q_k}\sum_{p\in\{\L,\N\}}\sum\limits_{\x_{mj}\in \Phi_{m,p}\setminus\{\x_{ki}\}}P_m h_{mj}\ell_{p}(x_{mj})G(\theta_{mj}).\label{eq:IntExpMain}
\end{align}
The noise power for  operator $k$ is given by $\Noise_k=N_0W_k$ where $N_0$ is the noise power density. Since $\Noise_k$ is dependent on the allocated bandwidth, it varies accordingly with association.  

\section{SINR and Rate Coverage Probability}\label{Sec:SINRRate}
One  metric that can be used to compare systems is the SINR coverage probability. It is defined as the probability that the SINR at the user from its associated BS is above a threshold $T$ 
\begin{align}
\Pc^{\mathrm{c}}(T)&=\prob{\SINR>T},
\end{align}
and is equivalently the  CCDF (complementary cumulative distribution function) of the SINR. In this section, we will first investigate the association of a typical user of $n^{th}$ network to a BS and then compute the coverage probability for this user. 
\subsection{Association Criterion and Probability}
Recall that a user of the $n^{th}$ operator can be associated with a BS of any operator from the set $S_n$. Let $E_{ki}$ denote the event that the typical user 
is associated with the BS $\BS_{ki}$ ({\em i.e.} the $i^{th}$ BS of operator $k$). Let us denote the distance of this BS by $x=x_{ki}$ and type by $s=s_{ki}$ for compactness. The  event $E_{ki}$ is equivalent to the event that no other BS  has higher $P^{\mathrm{avg}}$ at the user. This event can be further written as combination of following two events:  (i) the event that no other BS of operator $k$ has higher  $P^{\mathrm{avg}}$ at the user, and (ii) that no BS of any other accessible operator  $m$ has higher  $P^{\mathrm{avg}}$ at the user:
\begin{align}
E_{ki}
&=\{P^{\mathrm{avg}}_{ki}> P^{\mathrm{avg}}_{kj} \ \forall i\ne j \}
\iftoggle{SC}{}{\nonumber\\&} \cap
\{    P^{\mathrm{avg}}_{ki}> P^{\mathrm{avg}}_{mj} \ \forall  m\in S_n\setminus\{k\} \}\label{eq:Eki1}.
    \end{align}
Substituting  \eqref{eq:avgrxpower} in \eqref{eq:Eki1}, 
    \begin{align*}
E_{ki}&=\left\{\frac{C_\losind }{(x_{ki})^{\a_\losind}} >
    \frac{C_{s_{kj}}}{ (x_{kj})^{\a_{s_{kj}}}}
    \ \forall i\ne j \right\} \bigcap
\iftoggle{SC}{}{\\& \equalspace }
\left\{\frac{C_\losind P_k}{ (x_{ki})^{\a_\losind}}
   >
   \frac{C_{s_{mj}}P_m}{ (x_{mj})^{\a_{s_{mj}}}}\ \forall   
 m\in S_n\setminus\{k\} \right\}.
 \end{align*}
 The above condition can also be further split over LOS and NLOS tiers of each network, then it can be expressed as an equivalent condition over locations of all BSs as follows:
 \begin{align}
 \iftoggle{SC}{}{&}
E_{ki}=  \iftoggle{SC}{&}{}\left\{x_{kj} > x\ \forall \  s_{kj}=s,i\ne j \right\} 
\iftoggle{SC}{}{\nonumber\\&}
\bigcap
    \left\{x_{kj} > \left(\frac{C_{\sco}}{C_s} 
    \right)^{\frac{1}{\a_{\sco}}}x^{\frac{\a_s}{\a_{\sco}}}
    \ \forall   s_{kj}\ne s,i\ne j\right\} \nonumber
    \\ &  \bigcap \left\{ x_{mj} > 
    \left( \frac{P_m}{P_k} \right)^{\frac{1}{\a_{s}}} x
    \ \forall s_{mj}=s, m \in S_n\setminus \{k\} \right\}
    \nonumber \\
 &\bigcap\left\{x_{mj} > 
     \left(\frac{P_mC_{\sco}}{P_kC_s}   
     \right)^{\frac{1}{\a_{\sco}}}
     x^{\frac{\a_s}{\a_{\sco}}}\ 
     \forall  s_{mj}\ne s,  m\in S_n\setminus \{k\} \right\}
     \nonumber
\end{align}
where $\sco$ denotes the complement of the link type $s$.  {In other words, if $s=\L$, then $\sco=\N$ and if $s=\N$, then $\sco=\L$. 
The first condition restricts all BSs of operator $k$ with link type $s$ (same as type of serving BS) to be located outside a 2D ball. The second term is for all BSs of operator $k$ and link type $\sco$. Similarly the third and fourth terms are for BSs of all other accessible operators with link type $s$ and $\sco$, respectively.} 

As seen from these conditions, the average received power based association rule  effectively creates exclusion regions around the user for BSs of each network in $S_n$. 
Let us denote the exclusion radius for the tier $\{m,p\}$ by $D^{ks}_{mp}(x)$. For example, the exclusion region for all  LOS BSs of  operator $m$ when the user is associated with a NLOS BS of operator $k$ is given by
\begin{align}
D^{k\N}_{m\L}(x)=\left(\frac{P_m}{P_k} \frac{C_{\L}}{C_{\N}}    
     \right)^{\frac{1}{\a_{\L}}}x^{\frac{\a_\N}{\a_\L}}.
\end{align}
Note that for  BSs of the networks that are not in set $S_n$, there are no exclusion regions, \ie $D^{ks}_{mp}(x)=0\ \forall m\notin S_n$. {This exclusion region denotes the region where interfering BSs cannot be located and hence, affects the sum interference. }

The probability that all BSs of the tier $\{m,p\}$ are outside the exclusion radius $d$ is given by the void probability of the PPP $\Phi_{mp}$ which is $\VoidF{m,p}{d}= \exp(-\Lambda_{m,p}(\Ball(d)))$. 
Since the PPPs of the tiers are mutually independent, the probability that  BSs of the tiers other than $\{k,i\}$ are located outside the exclusion region, can be calculated by multiplying the individual void probabilities of each tier:
\begin{align}
f^o_{k,s}(x)&= \VoidF{k,\sco}{
      D^{ks}_{k\sco}(x) }
      \iftoggle{SC}{}{\nonumber\\&}
      \prod_{m\in S_n\setminus \{k\} } 
       \VoidF{m,s}{D^{ks}_{ms}(x)}
    \VoidF{m,\sco}{ D^{ks}_{m\sco}(x)} .
\end{align}
 Therefore, the probability density function of the distance $x$ to this associated BS is given as
\begin{align}
f_{k,s}(x)
&=  2\pi\lambda_{k,s}  x  \ 
	\VoidF{k,s}{x}\ 
	\VoidF{k,\sco}{
      D^{ks}_{k\sco}(x) }
       \iftoggle{SC}{}{\nonumber\\&}
      \prod_{m\in S_n\setminus \{k\} } 
       \VoidF{m,s}{D^{ks}_{ms}(x)}
    \VoidF{m,\sco}{ D^{ks}_{m\sco}(x)}.
  \label{eq:AssocProbExp}
\end{align}    
The probability that a user of operator $n$ is associated with a BS of operator $k$ can be computed by summation over both LOS and NLOS tiers: 
\begin{align}
A^n_k&=\int_0^{\infty}\left(f_{k,\L}(x)+f_{k,\N}(x)\right)\intd x.
\end{align}

 Let $\Pc^\mathrm{c}_{k\L}$ and $\Pc^\mathrm{c}_{k\N}$ denote the coverage probabilities for the typical user which is associated with a LOS and NLOS BS of operator $k$, respectively. They can be computed by integrating the CCDF of $\SINR$ from serving BS over pdf of distance $x$ from serving BS as follows:
\begin{align}
 \iftoggle{SC}{}{&}
\Pc^\mathrm{c}_{ks}(T) \iftoggle{SC}{&}{}=\int_0^\infty \prob{\SINR_{ks}(x)>T}f_{k,\L}(x)\intd x\nonumber\\
&=\int_0^\infty \prob{P_kh_{ks}C_{s}G(0)>T(I+\Noise_k)x^{\a_{s}}}f_{k,s}(x)\intd x.\label{eq:pcstep1}
\end{align}
 Since $h_{ks}\sim\exp(1)$, the probability in \eqref{eq:pcstep1} can be  replaced as
 \begin{align}
  \iftoggle{SC}{}{&}
\Pc^\mathrm{c}_{ks}(T)
 \iftoggle{SC}{&}{}
 =\int_0^\infty 
        \expect{\exp{\left(-\frac{T\Noise_kx^{\a_{s}}}{C_s G_1P_k}-
        \frac{TIx^{\a_s}}{C_s G_1P_k}\right)}}
        f_{k,s}(x)\intd x\nonumber\\
&=\int_0^\infty 
        \exp{\left(-\frac{TN_0W_kx^{\a_s}}{C_\L G_1P_k}\right)}
        \laplace{I}\left(\frac{Tx^{\a_s}}{C_s G_1P_k}\right)
        f_{k,\L}(x)\intd x
\end{align}
where $\laplace{I}(t)$ denotes the Laplace transform of the interference $I$ caused  by BSs of all networks in set $Q_k$ and is defined as $ 
\laplace{I}(t)=\expect{e^{-tI}}$.


Since the association with different tiers are disjoint events, the SINR coverage probability of the typical user can be computed by summing these individual tier coverage probabilities over all accessible tiers:
\begin{align}
\Pc^{\mathrm{c}}(T)&=\sum_{k\in S_n}\Pc^\mathrm{c}_{k}(T)=\sum_{k\in S_n}\left[\Pc_{k\L}^\mathrm{c}(T)+\Pc_{k\N}^\mathrm{c}(T)\right], 
\label{eq:CovProbBase1}
\end{align}
where $\Pc_k^\mathrm{c}(T)$ is the sum  of coverage probabilities of both tiers of the operator $k$.
To proceed further, we need to first characterize the interference $I$ \addvzerofour{for which we will compute its Laplace transform  defined as $\laplace{I}(t)=\expect{e^{-tI}}$}.

 \subsection{Interference Characterization}
If the user of operator $n$ is associated with operator  $k$, it experiences interference from all networks operating in spectrum $W_k$. Recall that all these interfering networks form set $Q_k$. Hence, the total interference is given by \eqref{eq:IntExpMain}.
{Due to mutual independence of the tiers, its Laplace transform can written as product of the following terms:
\begin{align}
\laplace{I}(t)&=\expect{e^{-tI}}=\prod_{m\in Q_k}{\laplace{I_{m}}(t)}=\prod_{m\in Q_k}\left({\laplace{I_{m\L}}(t)}{\laplace{I_{m\N}}(t)}\right)
\end{align}
 where $\laplace{I_{m}}(t)$ refers to the interference caused by network $m$ and $\laplace{I_{m\L}}(t)$ and $\laplace{I_{m\N}}(t)$ denote the Laplace transforms of LOS and NLOS interference from network $m$ }which are given in the following Lemma. 

\begin{lemma}\label{lemma:laplace}
The Laplace transforms of the interference  from LOS and NLOS BSs of network $m$ to a user of operator $n$ which is associated with a type $s$ BS of operator $k$ in a multi-operator  system are given as
 \begin{align*}
 \laplace{I_{m\L}}(t)
&
=\exp{}
        \left(
            -2\lambda_{m}[
            \thetab
            F_\L(\beta,\a_\L,tG_1P_mC_\L,D^{ks}_{m\L}(x))
          \iftoggle{SC}{}{  \right.\\ & \equalspace\equalspace \left.}
            +(\pi-\thetab)
            F_\L(\beta,\a_\L,tG_2P_mC_\L,D^{ks}_{m\L}(x))
            ]
        \right)\\
 \laplace{I_{m\N}}(t)
 &=\exp{}
        \left(
            -2\lambda_{m}[
            \thetab
            F_\N(\beta,\a_\N,tG_1P_mC_\N,D^{ks}_{m\N}(x))
            \iftoggle{SC}{}{\right.\\ & \equalspace\equalspace \left.}
            +(\pi-\thetab)
            F_\N(\beta,\a_\N,tG_2P_mC_\N,D^{ks}_{m\N}(x))
            ]
        \right)
 \end{align*}
 \begin{align*}
 \text{where }
 F_\L(b,a,A,x)&=\int_x^{\infty} \frac{e^{-by}Ay^{-a}}{1+Ay^{-a}}  y \intd y,
 \text{ and } 
  \iftoggle{SC}{}{\\}
  F_\N(b,a,A,x)&=\int_x^{\infty}(1-e^{-by}) \frac{Ay^{-a}}{1+Ay^{-a}}  y \intd y.
 \end{align*}
\end{lemma}
\begin{IEEEproof}
See Appendix \ref{lemmaproof:laplace}
\end{IEEEproof}

Note that the term containing  $G_1$ and $\thetab$ denotes the interference from the aligned BSs whose antennas are directed towards the considered user while the term containing $G_2$ and $(\pi-\thetab)$ represents the interference from the  unaligned BSs. Since $G_1 \gg G_2$, the interference from the aligned BS is significantly larger than the interference from the unaligned BS and hence, dominates the Laplace transform expression. Therefore, the value of  $\thetab$ plays a significant role in characterizing the interference and determining the benefits of spectrum license sharing. 

Now, we  provide the final expression for SINR coverage probability. 

\begin{theorem}\label{thm:CovProb}
The   SINR coverage probability of a typical user of operator  $n$ in a multi-operator mmWave cellular system is given as 
\begin{align}
\Pc^\mathrm{c}(T)=
\sum_{\substack{k\in S_n}}\sum_{s\in \{\L,\N\}}\int_0^\infty
\prod_{m\in Q_k}\laplace{I_{m\L}}\left(\frac{Tx^{\a_s}}{C_s G_1P_k}\right) 
\iftoggle{SC}{}{\nonumber\\}
\laplace{I_{m\N}}\left(\frac{Tx^{\a_s}}{C_s G_1P_k}\right)
\exp{\left(-\frac{N_0W_kTx^{\a_s}}
{C_s G_1P_k}\right)}
f_{k,s}(x)\intd x\label{SINRExp}
\end{align}
where $\laplace{I_{mp}}(t)$ is computed in Lemma \ref{lemma:laplace} and $f_{k,s}(x)$ is given in \eqref{eq:AssocProbExp}.
\end{theorem}
\begin{IEEEproof}
Substituting the value of $\laplace{I}(t)$ from Lemma \ref{lemma:laplace} in \eqref{eq:CovProbBase1}, we get the result.
\end{IEEEproof}

In \eqref{SINRExp}, the first summation  is over all networks which the user of operator $n$ can connect to, weighted by the association probability. This weighting is included inside the term $f_{k,s}(x)$. 

\addvone{Since due to complexity of the above expressions, it is difficult  to derive  direct insights, we also consider a simple case to simplify the expressions.
\begin{corollary}\label{cor:simplecase}
Consider a mmWave system with $n$ identical operators with closed access and fully shared licenses. Assuming that the LOS and NLOS channel have same pathloss parameters with $\alpha_\L=\alpha_\N=4$ and the side lobe gain $G_2=0$, the probability of coverage for  the interference limited scenario is given as
\begin{align}
\Pc_\mathrm{c}(T)=\frac{1}{1+\frac{\thetab}{\pi}T^\frac{1}{2}\left(n\frac\pi2-\arctan(T^{-\frac12})\right)}\label{eq:simplePc}.
\end{align}
\end{corollary}
It can be observed from the \eqref{eq:simplePc} that the probability of coverage decreases monotonically with the number of operators $n$.}

\subsection{Rate Coverage}
\addvone{While the SINR  shows the serving link quality, the per-user rate  represents the data bits received per second per user and is one of the main goal for using mmWave bands.} In this section, we derive the downlink rate coverage which is defined as the probability of the rate of a typical user being greater than the threshold $\rthres$,
\begin{align}
\Rc(\rthres)&=\prob{\mathrm{Rate}>\rthres}.
\end{align}

Let us assume that $O_k$ denotes the time-frequency resources allocated to each user associated with the `tagged' BS of operator $k$. Therefore, the instantaneous rate of the considered typical user is given as $R_{ki}=O_k\log{\left(1+\SINR_{ki}\right)}$.
 The value of $O_k$  depends upon the
number of users ($N_k^\user$), equivalently the load, served by the tagged BS.
The load  $N_k^\user$ is a random variable due to the randomly sized coverage areas of each BS and random number of users in the coverage areas. As shown in \cite{SinDhiJ2013,SinghBackHaul2015} approximating this load with its respective mean does not compromise the accuracy of results. 
Since  user distribution of each network is assumed to be PPP, the average number of users associated with the tagged  BS of network $k$ associated with the typical user can be modeled similarly to \cite{SinDhiJ2013,SinghBackHaul2015}:
\begin{align}
N^\user_k&=1+1.28\frac1{\lambda_k}\sum_{m:k \in S_m}\lambda^\user_m{A^m_k}.
\end{align}
Note
that the summation is over all the networks whose users can connect to the network $k$ and the sum denotes the combined density of associated users from each network.  

{Now, we assume that the scheduler at the tagged BS gives  $1/N_k^\user$ fraction of resources to each of the $N_k^\user$ users. This assumption can be justified as most schedulers such as round robin or proportional fair  give  approximately  $1/N_k^\user$ fraction of resources to each user on average.}
Using the mean load approximation, the instantaneous rate of a typical user of operator $n$ which is associated with $i^{th}$ BS of operator $k$ is given as
\begin{align}
R_{ki}&=\frac{W_k}{N^\user_k}\log{\left(1+\SINR_{ki}\right)}\label{eq:InRateExp}.
\end{align} 

Let $\Rc^\mathrm{c}_k(\rthres)$ denote the rate coverage probability when user is associated with  operator  $k$. Then the total rate coverage will be equal to the sum of $\Rc^\mathrm{c}_k(\rthres)$'s over all accessible networks:
\begin{align}
\Rc^\mathrm{c}(\rthres)&=\sum_{k\in S_n} \Rc^\mathrm{c}_k(\rthres)\label{eq:RateExp}.
\end{align}
$\Rc^\mathrm{c}_k(\rthres)$ can be derived in terms of SINR coverage probability as follows:
\begin{align*}
\Rc^\mathrm{c}_k(\rthres)&=\prob{R_{ki}>\rthres}=\prob{W_k/N^\user_k\log{(1+\SINR_{ki})}>\rthres}\\
&=\prob{\SINR_{ki}>2^{\rthres\frac{N^\user_k}{W_k}}-1}=\Pc_k^\mathrm{c}\left(2^{\rthres N_k^\user/W_k}-1\right).
\end{align*}
Now, the rate coverage is given as
\begin{align}
\Rc^\mathrm{c}(\rthres)&=\sum_{k\in S_n} \Pc_k^\mathrm{c}\left(2^{\rthres N_k^\user/W_k}-1\right)\label{eq:FinalRateExp}.
\end{align}

\addvone{We, now, turn back to the simple case considered in  Corollary \ref{cor:simplecase} to simplify the expressions and provide further insights.
\begin{corollary}\label{cor:simplecase2}
Consider a mmWave system with $n$ identical operators with closed access and fully shared licenses of bandwidth $\bar{B}$ each as considered in  Corollary \ref{cor:simplecase}. The rate coverage for  the interference limited scenario is given as
\begin{align}
\Rc_\mathrm{c}(\rthres)=
\frac{1}
{1+\frac{\thetab}{\pi}
	\left(2^{\rthres'/n}-1\right)^{\frac12}
	\left(n\frac\pi2-\arctan((2^{\rthres'/n }-1)^{-\frac12})\right)}\label{eq:simpleRc}
\end{align}
where $\rthres'={\rthres N^\user}/{\bar{B}}$.
\end{corollary}
The denominator in \eqref{eq:simpleRc} behaves differently with respect to $n$ for different regimes of $\rthres'$. For small $\rthres$,  rate coverage decreases with respect to $n$ making exclusive licenses more beneficial. For large  $\rthres$, the rate coverage increases with respect to $n$ which favors sharing spectrum licenses.   Let us compare the case of $n=1$ and $n=2$. For $\thetab=\pi$, the license sharing becomes more beneficial than exclusive licenses when $\rthres'>3.3$ which is equivalent to  $\Rc_\mathrm{c}=0.20$  indicating that at maximum, only 20\% users would be benefiting if the licenses are shared. This effectively means that spectrum sharing is not beneficial from the operator's perspective. Whereas for $\thetab=10^\mathrm{o}$, the switch occurs at $\Rc_\mathrm{c}=0.83$  indicating that 83\% users are benefiting from spectrum sharing and making sharing of licenses favorable from the operator's perspective. Therefore, the directionality of antennas ($\thetab$) plays a crucial role in determining the  least value of $\Rc_\mathrm{c}$ where  license sharing starts becoming more beneficial than exclusive licenses. We later show  similar trends with target rate and beamwidth using simulations.
}


\section{Performance Comparison}\label{Sec:4}
We use the preceding mathematical framework to compare the benefits of spectrum license sharing. 
We enumerate three specific cases (or systems) considering different combinations of accesses and  license sharing schemes. Also see \figurekeyword \ref{fig:sysmodel} for a visual explanation.


\textbf{System 1: Closed Access and Exclusive Licenses:} 
In System 1, each user must associate with only its own network  and each operator can use its own spectrum only.   This case is equivalent to a set of $\K$ single operator systems which has been studied in prior work \cite{Bai2014}. This system serves as a baseline case to evaluate benefits of sharing.  
The SINR coverage probability of \UE is given  by \eqref{SINRExp} with $k=n$, $S_n=\{n\}$. 
 Recall that in this system, the spectrum accessible to each operator is  its own licensed spectrum only, \ie $W_n=B_n$, $Q_n=\{n\}$.

\textbf{System 2: Open Access and Full Spectrum License Sharing: }
In System 2, each user can be associated with any network  and the spectrum license is shared between all operators. 
In this case, the probability of association of a user with any BS of  the operator  $k$ is independent of which operator this user belongs to and is given as
\begin{align*}
A^n_{k}&=A_k=A_{k,\L}+A_{k,\N}=\int_0^{\infty}\left(f_{k,\L}(x)+f_{k,\N}(x)\right)\intd x.
\end{align*}
The SINR coverage probability of \UE is given  by \eqref{SINRExp} with $S_n=\{1,2,\cdots \K\}$. 
The average load to the tagged BS of operator $k$ is given as
$N^\user_k=1+1.28\sum_m{\lambda^\user_m}\frac{A_k}{\lambda_k}$. The spectrum accessible to each operator $W_k$  is the complete band $B$ and $Q_n=\{1,2\cdots\K\}$. 
Since users can be served by BS of any operator, it requires full coordination, sharing of control channels and technology sharing among operators. The quality of service to a user will be the same regardless of the operator it belongs to. This will limit the technological advantage of an operator over other operators and hence, operators may not want to open up their networks to each other. Therefore, System 2 is likely not a practical system but instead serves as an upper bound to the  two other more practical systems.
Note that if all $\K$ networks are identical  with respect to every parameter, then this system is equivalent to a single operator system with the aggregate BS and UE density.  \addvzerofour{In this case,} from a user association perspective, there is no discrimination based on the network which a particular BS belongs to. Also all networks transmit in the same band. So the users effectively see a single network with aggregate BS density of all the networks. {Similarly from the operator's perspective, users of all the networks look the same due to open access. Hence, the users of different operators can be replaced by users of a single operator with  the aggregate UE density. }

\iftoggle{SC}{
\begin{figure}[!t]
\centering
\includefigtrimmed{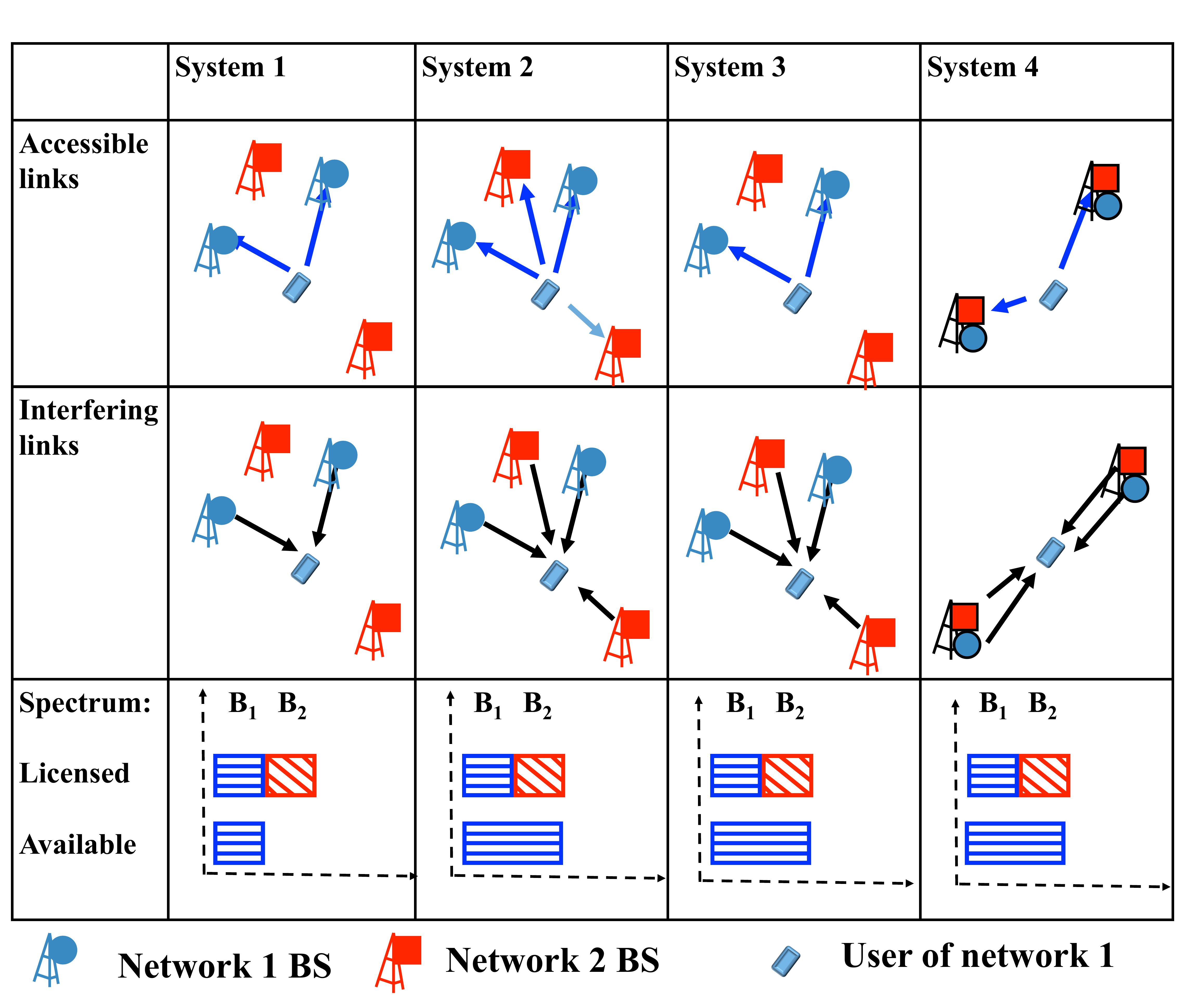}{0 10 0 40}
\caption{Illustration describing the differences between the four  systems. For a typical user of operator 1, the figure shows all accessible networks this user can connect to, all interfering networks, and the available spectrum after license sharing. }
\label{fig:sysmodel}
\end{figure}}
{
\begin{figure*}[!t]
\centering
\includefigtrimmed{SYSMODV0P6}{0 10 0 40}
\caption{Illustration describing the differences between the four  systems. For a typical user of operator 1, the figure shows all accessible networks this user can connect to, all interfering networks, and the available spectrum after license sharing. }
\label{fig:sysmodel}
\end{figure*}
}

\textbf{System 3: Closed Access and Full Spectrum License Sharing:} 
In System 3, each user must associate with its own network but the whole spectrum is shared between all the operators. {This case does  not require any transmission coordination  among networks or common control channels, {nor  does it} require sharing of infrastructure or back-haul resources. This system is close to the practical implementation where subscribers must connect to their respective service providers only.} 
The SINR coverage probability of the considered typical user of operator $n$ is given by \eqref{SINRExp} with $k=n$, $S_n=\{n\}$. 
For this system, the spectrum accessible to each operator is $B$ and $Q_n=\{1,2\cdots\K\}$.

Along with the above three systems, we consider  one additional system System 4 as defined below. This system will help us understand if independent operators can still share BS tower infrastructure while sharing the spectrum licenses. As mentioned before, there may be correlation among BSs locations of different operators with possible co-location of BSs, therefore it is important to understand how gain from license sharing will be affected when the correlation is present.

\textbf{System 4: Co-located BSs with Closed Access and Full Spectrum License Sharing:} 
System 4 has closed access and fully shared licenses where the respective BSs of all the operators are all co-located. The system model for  this case remains the same as the previous three systems except for the following two differences: 1) The BS locations are modeled by a single PPP $\Phi=\{x_j\}$ with intensity $\lambda$ and 2)  for a typical user, the BSs of all the operators located at the same location are either all LOS or all NLOS.
 We first briefly show the computation the probability of SINR coverage of this system. Consider a typical user of operator $n$. 
 The BS PPP $\Phi$ can be divided into two independent PPPs, $\Phi_\L$ and $\Phi_\N$ with intensity $\lambda_\L(x)=\lambda p(x)$ and $\lambda_\N(x)=\lambda (1-p(x))$. The probability density function of the distance $x$ to the associated BS of operator  $n$ is given as
\begin{align}
f_{s}(x)
&=  2\lambda_{s} \pi x 
    \exp\left(-\Lambda_{s}(\Ball(x))\right)
\iftoggle{SC}{}{\nonumber    \\&    \equalspace }
\exp\left(-\Lambda_{\sco}\left(
    \Ball \left(
      D^{s}_{\sco} (x)\right)
      \right)\right) \label{eq:AssocProbExpcolocs}
\end{align}
where the exclusion radius $D^{s}_{p}(x)$ is the same for BSs of all the operators and given as 
$D^{s}_{p}(x)={\left(\frac{C_p}{C_s}\right)}^{\frac1{\alpha_p}}x^{\frac{\alpha_s}{\alpha_p}}$.
The interference $I$ at the user from BSs of all the operators is given as
\begin{align}
I&=x^{-\alpha_s}C_s\sum_{m\in Q_n\setminus\{n\}}P_m h_{mi} G(\theta_{mi})
\iftoggle{SC}{}{\nonumber\\&}
+\sum_{p=\L,\N}\sum_{\x_j\in \Phi_p\setminus\{\x_i\}}x_j^{-\alpha_p}C_p\sum_{m\in Q_n}P_mh_{mj}G(\theta_{mj}).\label{eq:IntExpcolocs}
\end{align}

The following Lemma characterizes the Laplace transform of the interference in  \eqref{eq:IntExpcolocs} in the co-located BSs case.

\begin{lemma}\label{lemma:laplacecolocs}
The Laplace transform of interference to a typical user of operator $n$ with closed access which is associated to a BS of type $s$  in a multi-operator system with  co-located BSs is given as $\iftoggle{SC}{}{\laplace{I}(t)=}$
\begin{align*}
&\iftoggle{SC}{\laplace{I}(t)=}{}\prod_{{m\in Q_n\setminus\{k\}}}\left(\frac{\thetab/\pi}{1+tx^{-\alpha_s}C_sP_m G_1}+\frac{(\pi-\thetab)/\pi}{1+tx^{-\alpha_s}C_sP_m G_2}\right)\\
&\ \times\prod_{p=\L,\N}\exp{} \left(-2\pi\lambda\int_{D^s_p(x)}^\infty p(y)
\left({}1-\prod_{m\in Q_n}
\iftoggle{SC}{}{\right.\right.\\&\left.\left.}
{\left(\frac{\thetab/\pi}{1+ty^{-\alpha_p}C_pP_mG_1}+\frac{(\pi-\thetab)/\pi}{1+ty^{-\alpha_p}C_pP_mG_2}\right)}
{}
\iftoggle{SC}{}{\vphantom{\prod_{m\in Q_n}}}
\right)y\intd y\right)
\end{align*}
\end{lemma}
\begin{IEEEproof}
See Appendix \ref{lemmaproof:laplacecolocs}.
\end{IEEEproof}

Similar to previous subsections, the coverage probability of the considered typical user is given as $\iftoggle{SC}{}{\Pc^\mathrm{c}(T)=}$
\begin{align}
\iftoggle{SC}{\Pc^\mathrm{c}(T)=}{}
&\sum_{s=\L,\N} \int_0^\infty 
        \exp{\left(-\frac{TN_0W_nx^{\a_s}}{C_s G(0)P_n}\right)}
        \laplace{I}\left(\frac{Tx^{\a_s}}{C_s G(0)P_n}\right)
        f_{s}(x)\intd x \label{colocsSINRProb}
\end{align}
where $ \laplace{I}(t)$ is given in Lemma \ref{lemma:laplacecolocs} and $f_s(x)$ is given in \eqref{eq:AssocProbExpcolocs}.
Since full sharing of license is assumed  for this system, the spectrum accessible to each operator is $B$ and and $Q_n=\{1,2\cdots\K\}$. Hence, similar to System 3, the average load is given as $N^\user_n=1+1.28\lambda^\user_n\frac1{\lambda_n}$.


\section{Partial Loading of  The Network}\label{Sec:PL}
In the previous section, we assumed that all the BSs have at least one user associated  and they are all transmitting. Such an assumption is justified when the user density is very high in comparison to the BS density resulting in many associated users per BS and negligible probability of any BS being inactive. For very dense networks, however, this assumption will  break down and many BSs will not be occupied at all times. An interesting case to consider is where the system is not fully loaded and there are some BSs having no (active) users to associate with. In such a case, interference will  reduce which should favor license sharing. From \cite{SinDhiJ2013}, the number of users associated with a BS of network $m$ for a closed access system can be approximated as the following distribution:
\begin{align}
\kappa_m(N^\user_m)=\frac{3.5^{3.5}}{\Gamma(n+1)}\frac{\Gamma(n+3.5)}{\Gamma(3.5)}{\left(
\eta_m
\right)
}^n{\left(3.5+
\eta_m
\right)}^{-n-3.5}
\end{align}
where $\eta_m$ denotes the ratio between density of associated users and BS density for network $m$. The $\kappa_m(N^\user_m)$ approximation has been shown to closely match  the load distribution in  simulations  \cite{SinDhiJ2013,SinghBackHaul2015}. For the multi-network system, $\eta_m$ is computed as
\begin{align}
\eta_m&=\sum_{q:m \in S_{q}}\lambda^\user_{q}\frac{A^{q}_m}{\lambda_m},
\end{align}
where the sum is over all the networks whose users can connect to BSs of network $m$. The multiplication of association probability of a user of network $q$ and user density of network $q$ gives the density of the network $q$'s associated users for network $m$.

  The probability that a BS is off is equal to the probability that a BS has no user associated to it which is equal to $\kappa_m(0)$. Therefore for Systems 1, 2  and 3, the interfering BSs can be obtained by independent thinning of original BS PPP with probability $1-\kappa_m(0)$. The coverage probability of the Systems 1, 2 and 3 are given by Theorem \ref{thm:CovProb} with  $\lambda_m$ substituted by $\lambda'_m=\lambda_m (1-\kappa_m(0))$.

For System 4, the sum interference $I$ at the user from BSs of all the networks in partial loading case is given as
\begin{align}
&I=x^{-\alpha_s}C_s\sum_{m\in Q_n\setminus\{n\}}P_m h_{im} G(\theta_{im})\delta_{mi}
\iftoggle{SC}{}{\nonumber\\&}
+\sum_{p=\L,\N}\sum_{\x_j\in \Phi_p\setminus\{\x_i\}}x_j^{-\alpha_p}C_p\sum_{m\in Q_n}P_mh_{mj}G(\theta_{mj})\delta_{mj}\label{eq:IntExpcolocsPL}
\end{align}
where $\delta_{mj}$ is an indicator which is 1 when BS $\BS_{mj}$ is on. Hence, the Laplace transform of interference in partial loading can be computed as (see Appendix \ref{appen:3} for proof)  $\iftoggle{SC}{}{\laplace{I}\left(t\right)}=$
\begin{align}
&\iftoggle{SC}{\laplace{I}\left(t\right)=}{}
\prod_{{m\in Q_k\setminus\{k\}}}\left(
\kappa(0)+(1-\kappa(0))\sum_{j=1,2}\frac{a_j/\pi}{1+
tx^{-\a_s}C_sP_m G_j}
\right)\nonumber\\
&\ \times \prod_{p=\L,\N}\exp{}\left(-2
\lambda\int_{D^s_p(x)}^\infty p(y)\left({}1-\prod_{m\in Q_k}
{}\left(\kappa(0)
\iftoggle{SC}{}{\right.\right.\right.\nonumber\\&\ \ \ \left.\left.\left.}
+(1-\kappa(0))
\sum_{j=1,2}\frac{a_j}{1+y^{-\alpha_p}{
t C_pP_mG_j}{}}
\right){}
{}\right)y\intd y\right){}\label{eq:laplacecolocsPL}
\end{align}
where  $a_1=\thetab,a_2=\pi-\thetab$. The coverage probability of this system is given by \eqref{colocsSINRProb}
with $ \laplace{I}(t)$  given in \eqref{eq:laplacecolocsPL} and $f_s(x)$  given in \eqref{eq:AssocProbExpcolocs}.

\section{Numerical Results}\label{Sec:Results}
In this section, we provide  results numerically computed from the analytic expressions derived in previous sections. We compare the four aforementioned systems to provide insights and discuss the impact of license sharing. For these numerical results, we consider a system consisting of two cellular operators with identical parameters, both operating in mmWave band. Each operator owns a network of BSs with  density of {$30/\text{km}^2$} which is equivalent to average cell radius of 103 m and have users with density of $200/\text{km}^2$.
We consider the exponential blockage model \ie $p(x)=\exp(-\beta x)$ with $\beta=0.007$ which has an average LOS region of 144 m. The transmit power is assumed to be 26dBm.
For most of the results, the operating frequency is 28GHz for which pathloss exponents for LOS and NLOS are $\a_{\L}=2,\a_{\N}=4$ and the corresponding gains are  $C_{\L}=-60\dB, C_{\N}=-70\dB$. The total system bandwidth is 200 MHz. We assume that each operator owns a license to 100 MHz. Recall that for System 1, each operator can use only its own spectrum. In Systems 2, 3 and 4, both operators share each other's spectrum licenses and therefore, get 200MHz of  spectrum. 

\begin{figure}[!t]
\centering
\includefig{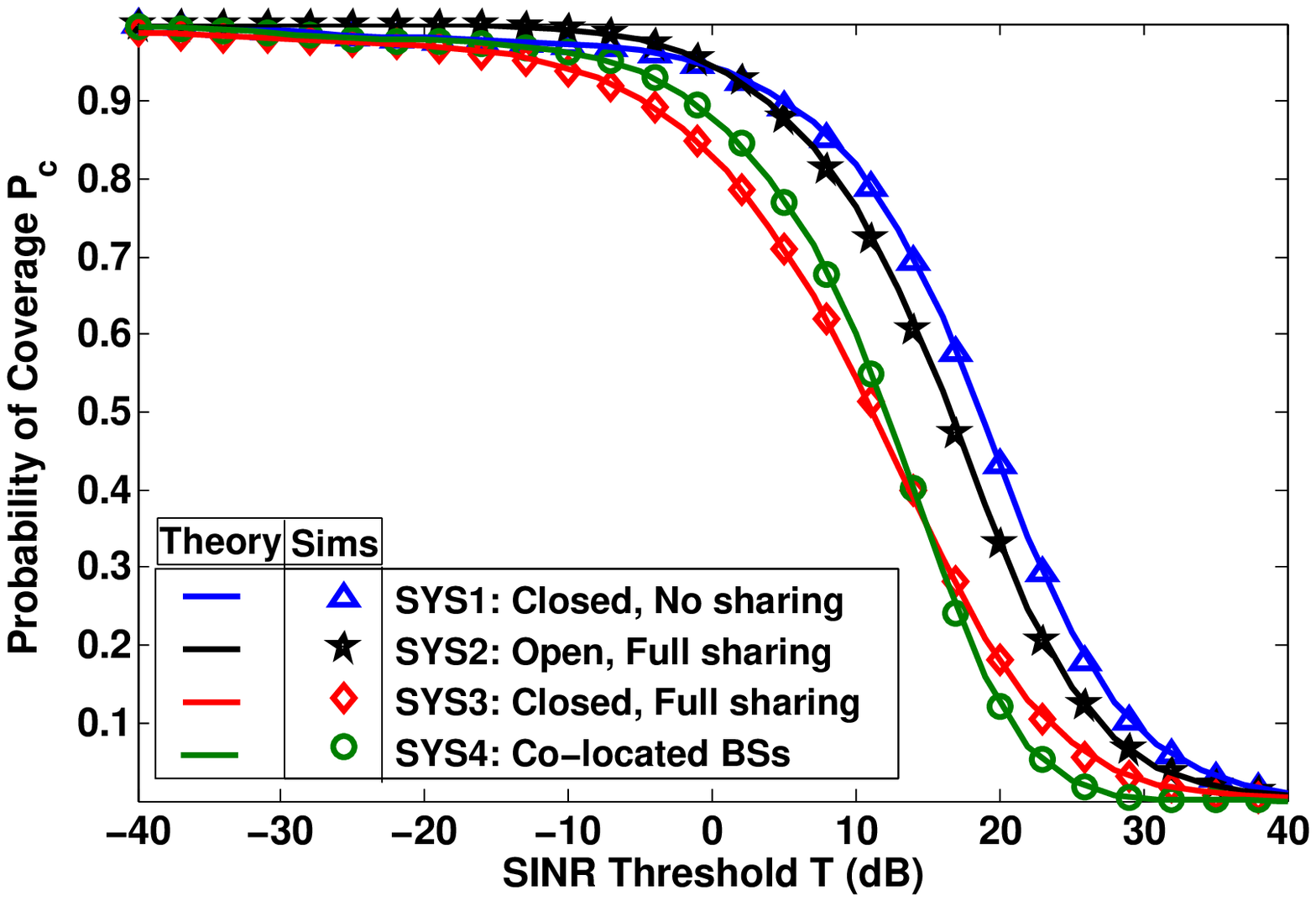}
\caption{Probability of SINR coverage in a two-network mmWave system with BS antenna half beamwidth $\thetab=10^{\mathrm{o}}$ for different cases. Line-curves denote values from the analysis and markers denote respective values from simulation.}
\label{fig:pc}
\end{figure}

\textbf{Validation of analysis and SINR coverage trends:} \figurekeyword \ref{fig:pc} compares the probability of coverage for these systems and validates our analysis with simulation.
The typical user  in System 2 has high SINR coverage due to its open access. 
The closed access in System 3 allows  BSs of another networks to be located closer than the serving BS and may lead to large interference. Therefore the user in System 3 has low SINR coverage. 
System 1 has the same closed access as System 3, but  the spectrum is not shared. Therefore the user faces no interference from other networks and hence, the SINR coverage is greater than System 3. In comparing System 1 and System 2, we observe different behaviors for different SINR ranges. Recall that System 2 is similar to System 1, but with double BS and MS density and double bandwidth. Hence, a serving BS is relatively closer in System 2 from System 1 by a factor of $\sqrt{2}$. Now, for the high SINR region (which is mainly due to LOS serving links
), this increases the received power of a serving BS by a factor of $\sqrt{2}^{\a_\L}=2$. Since the noise power also increases by a factor of 2 due to increase in bandwidth, it effectively cancels the increase in the power caused by increased proximity of serving BS. As far as interference is considered, since doubling density increases the probability of interferers to be  LOS,   interference increases significantly in System 2. Therefore SINR coverage is higher in System 1 than System 2. For the low SINR region (which is mostly due to NLOS serving links), the received power of a serving BS increases by a factor of $\sqrt{2}^{\a_\N}=4$ and the probability of serving link turning to LOS is also increased due to increased proximity of serving BS. Therefore System 2 has higher SINR coverage than System 1 in the low SINR region. System 4 has similar values and trends as System 3 when compared to other systems. \addvone{When we compare System 3 and System 4, we see a tradeoff  due to co-location of BSs. Since BSs are co-located in System 4, there cannot be any BS of other operators closer than the serving BS from the typical user. But in System 3, there can be an interfering BS closer to the user than the serving BS due to the independence assumption. This causes System 4 to perform better than System 3 for cell edge users where serving signal power is already low. But the same co-location argument also guarantees that System 4 will always have $\K-1$ interfering BSs of the other operators at the same location as serving BS, which is not the case in System 3. Therefore,  System 3 performs even better for the users with high serving signal power. Hence, we can observe  that SINR coverage of System 3 is better for high values of  SINR thresholds,  while System 4 performs better for low SINR thresholds.}

\begin{figure}[!t]
\centering
\includefig{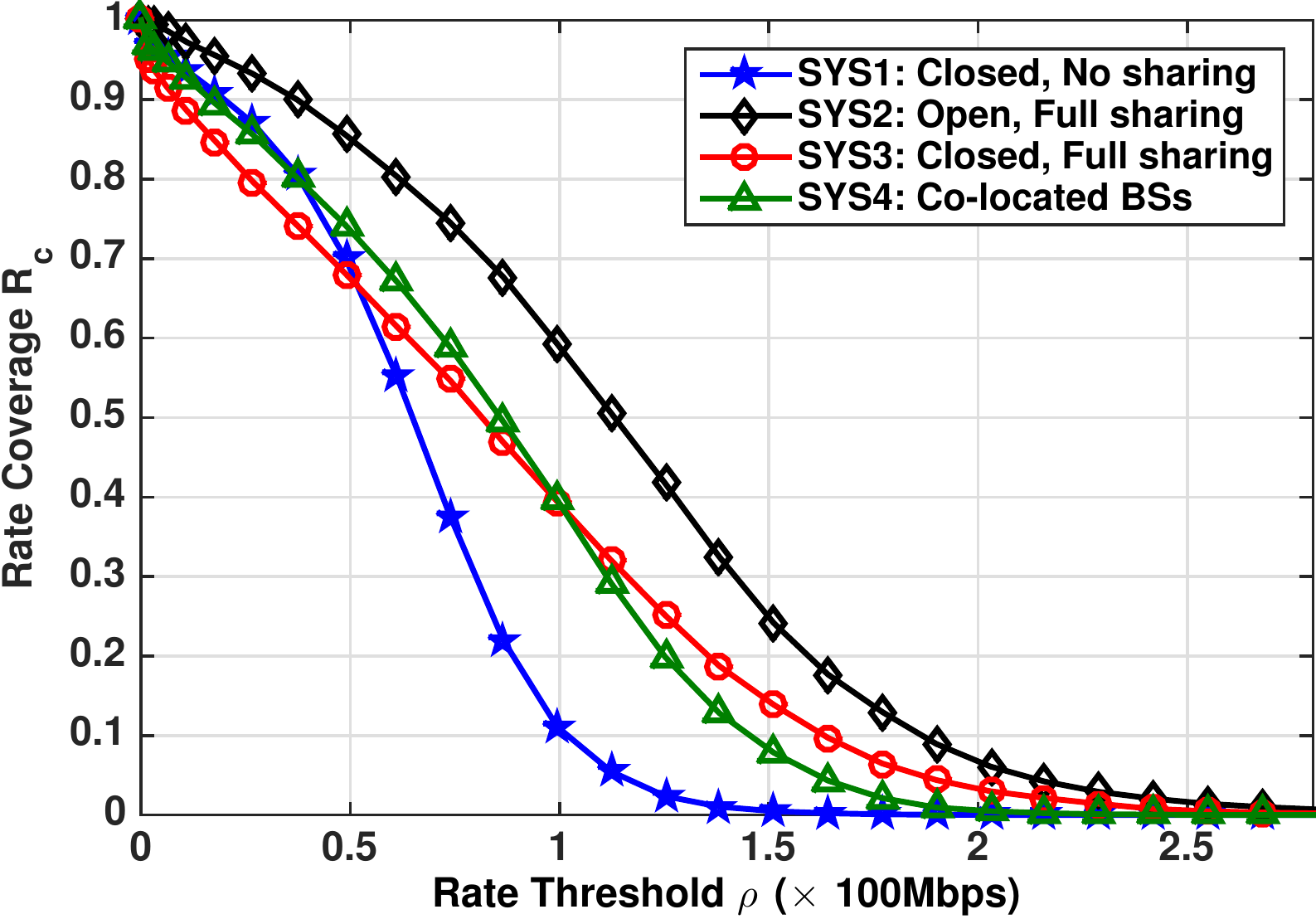}
\caption{Rate coverage in a two-network mmWave system  with Rayleigh fading and BS antenna half beamwidth $\thetab=10^{\mathrm{o}}$ for different cases. Systems 3 and 4 with shared license perform better than System 1 with exclusive licenses.}
\label{fig:rc}
\end{figure}

\textbf{Sharing licenses achieves higher rate coverage:} \figurekeyword  \ref{fig:rc}  compares the probability of rate coverage for four systems which incorporates the effect of load and bandwidth. Since each network has a large bandwidth and large SINR coverage in System 2, its rate coverage is the highest among all systems. Such a system, however, as mentioned earlier, may not be practical and  mainly serves as a benchmark  for practical systems. A more interesting comparison is between System 1 and 3 (or 4). Here we can see that even though System 1 has higher SINR coverage than System 3 (and 4), the latter achieves higher median rate, due to the extra bandwidth gained from spectrum license sharing. \addvzerofour{In particular, System 3 and 4 have respectively 25\% and 32\% higher median rates than System 1.}

\begin{figure}[!t]
\centering
\includefig{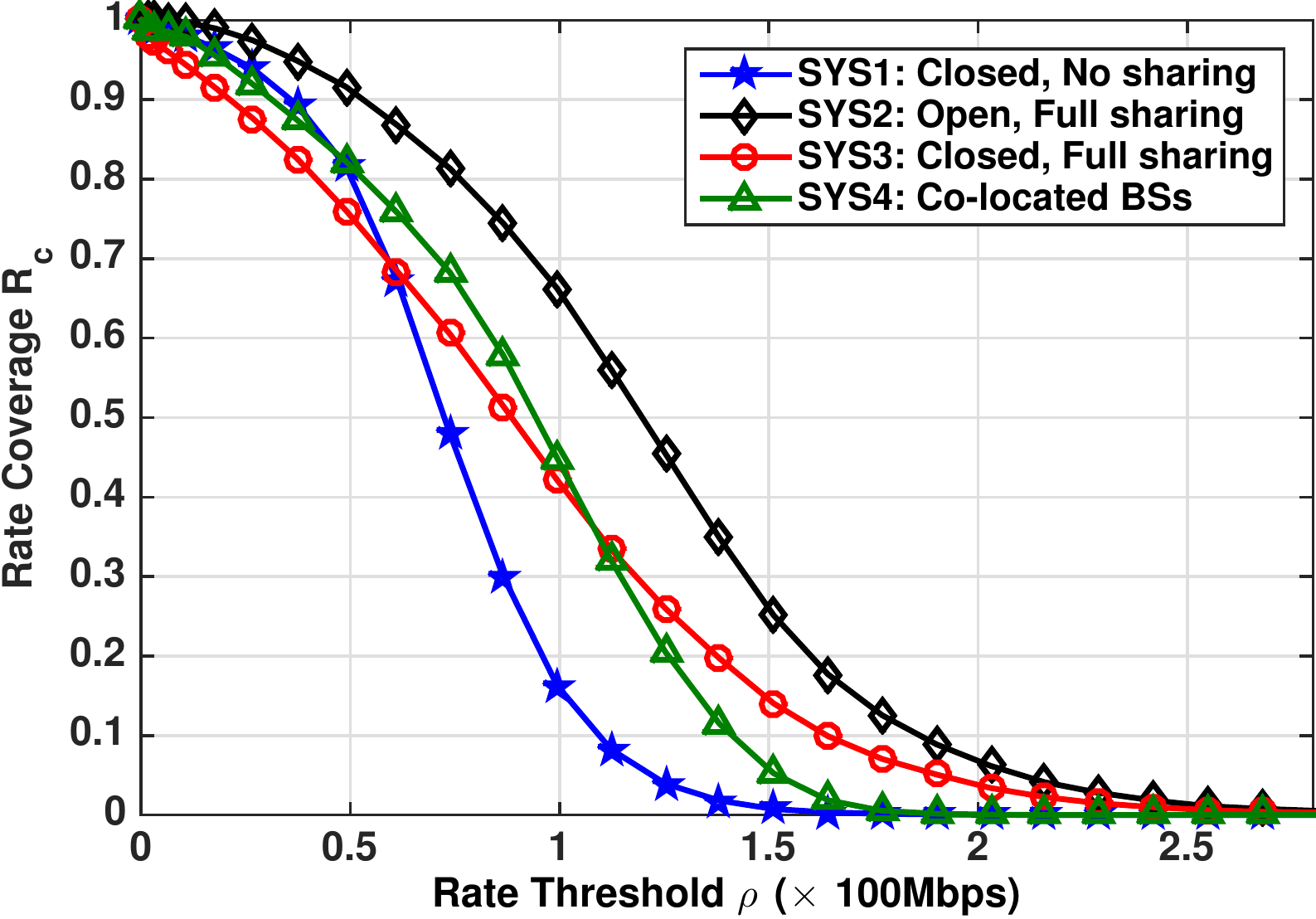} 
\caption{Rate coverage in a two-network mmWave system with grid BS deployment and 3GPP antenna pattern \cite{Antenna3GPP}  for different cases. \addvzerofour{The trends for this case are similar to Fig. \ref{fig:rc} which validates our assumptions regarding the system model}. }
\label{fig:rcreal}
\end{figure}

\begin{figure}[!t]
\centering
\includefig{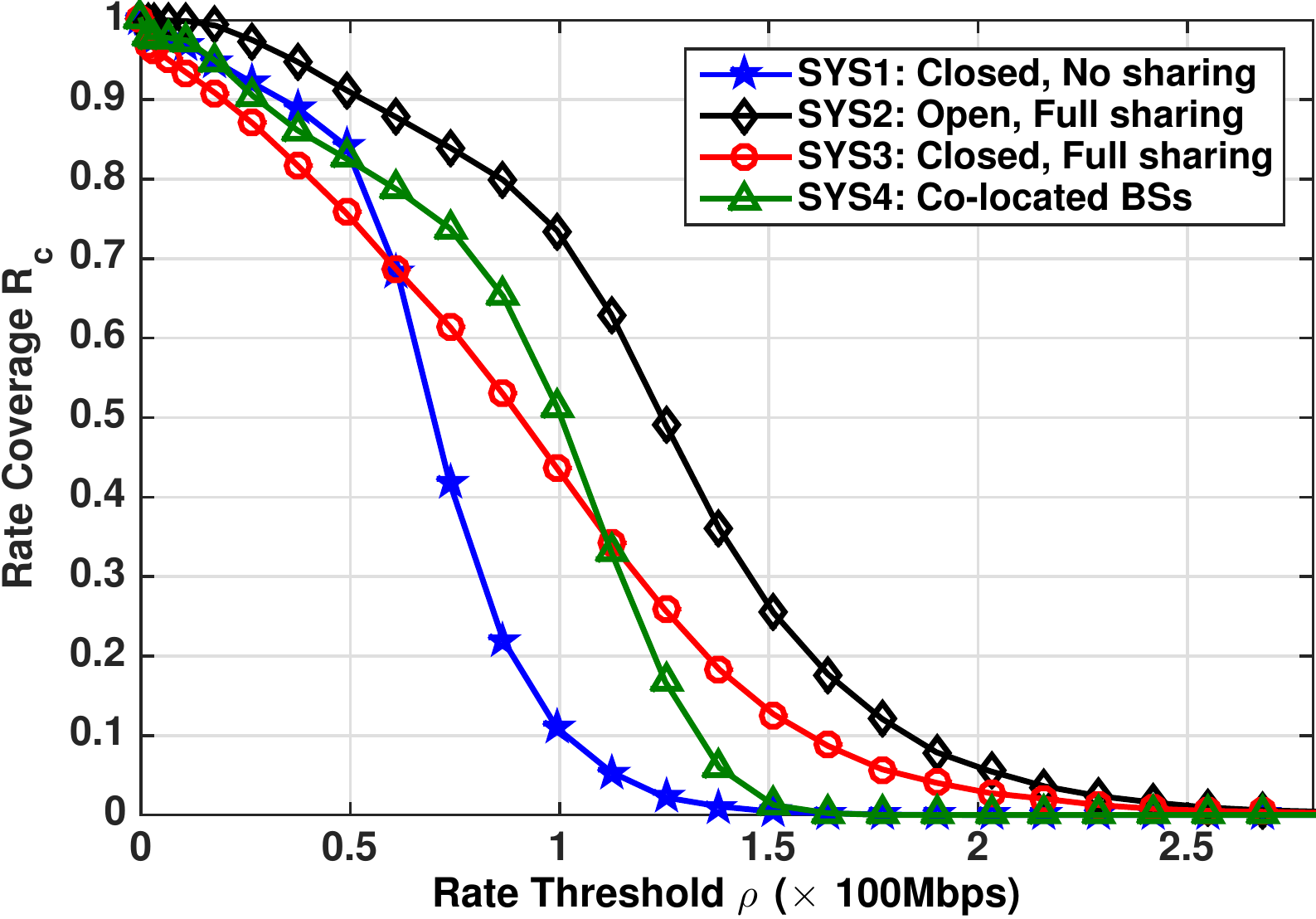}
\caption{Rate coverage in a two-network mmWave system  with Nakagami fading (with parameter 10) and BS antenna half beamwidth $\thetab=10^{\mathrm{o}}$ for different cases. When compared to Rayleigh fading (Fig. \ref{fig:rc}), the insights are similar which justifies the Rayleigh fading assumption for analysis}
\label{fig:fadingcompare}
\end{figure}

\textbf{Validation of the model with realistic scenario:} To validate our PPP assumption and to show that it is reasonable, we also present a simulation result  where the BSs are deployed in a square grid and the BS antenna pattern is parabolic, as specified by the 3GPP standard \cite{Antenna3GPP}.  \addvone{In System 2 and 3, both operators have their own square grid deployments which  is shifted  from each other by a random amount in each realization of  the simulation.} We also consider log-normal shadowing ($\sigma_{\mathrm{LOS}}=5.2$dB and $\sigma_{\mathrm{NLOS}}=7.6$dB). \figurekeyword \ref{fig:rcreal} compares the probability of rate coverage for four systems with these modifications. We observe similar trends for spectrum license sharing which justifies our assumptions regarding deployment and shadowing.
\figurekeyword \ref{fig:fadingcompare} shows the probability of rate coverage for four systems with Nakagami fading (with parameter $10$) instead of Rayleigh.  It can be seen that the trends are similar to those of Rayleigh distribution as we claimed in Section IIA. 


\begin{figure}[!t]
\centering
\includefig{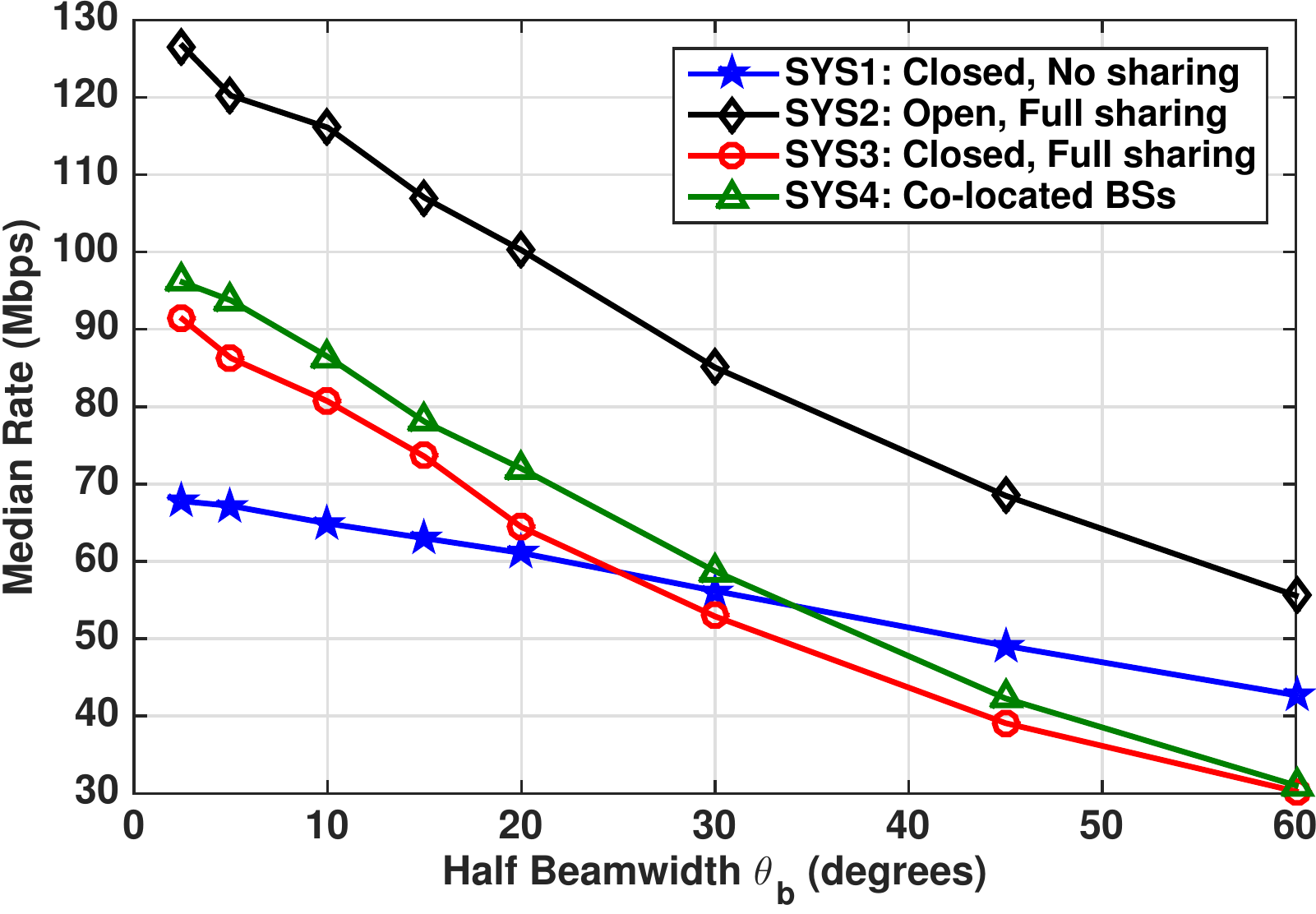}
\caption{Median rate versus BS antenna beamwidth in two-network mmWave system under different cases for Rayleigh fading. Systems with sharing of license outperforms System 1 with no shared license for moderate and low values of antenna beamwidth.}
\label{fig:mrctheta}
\end{figure}

\textbf{Impact of beamwidth on median rate:} \figurekeyword \ref{fig:mrctheta} compares the median rate of the four systems for various values of beamwidth. It can be seen that above a certain threshold for the beamwidth, it becomes more beneficial to have exclusive license, due to high interference. As the beamwidth decreases, license sharing becomes more beneficial. For the given parameters, the threshold is at about $25^\mathrm{o}$. Since mmWave has typical beamwidth less than $15^\mathrm{o}$, sharing should increase the achievable rate. 

\begin{figure}[!t]
\centering
\includefig{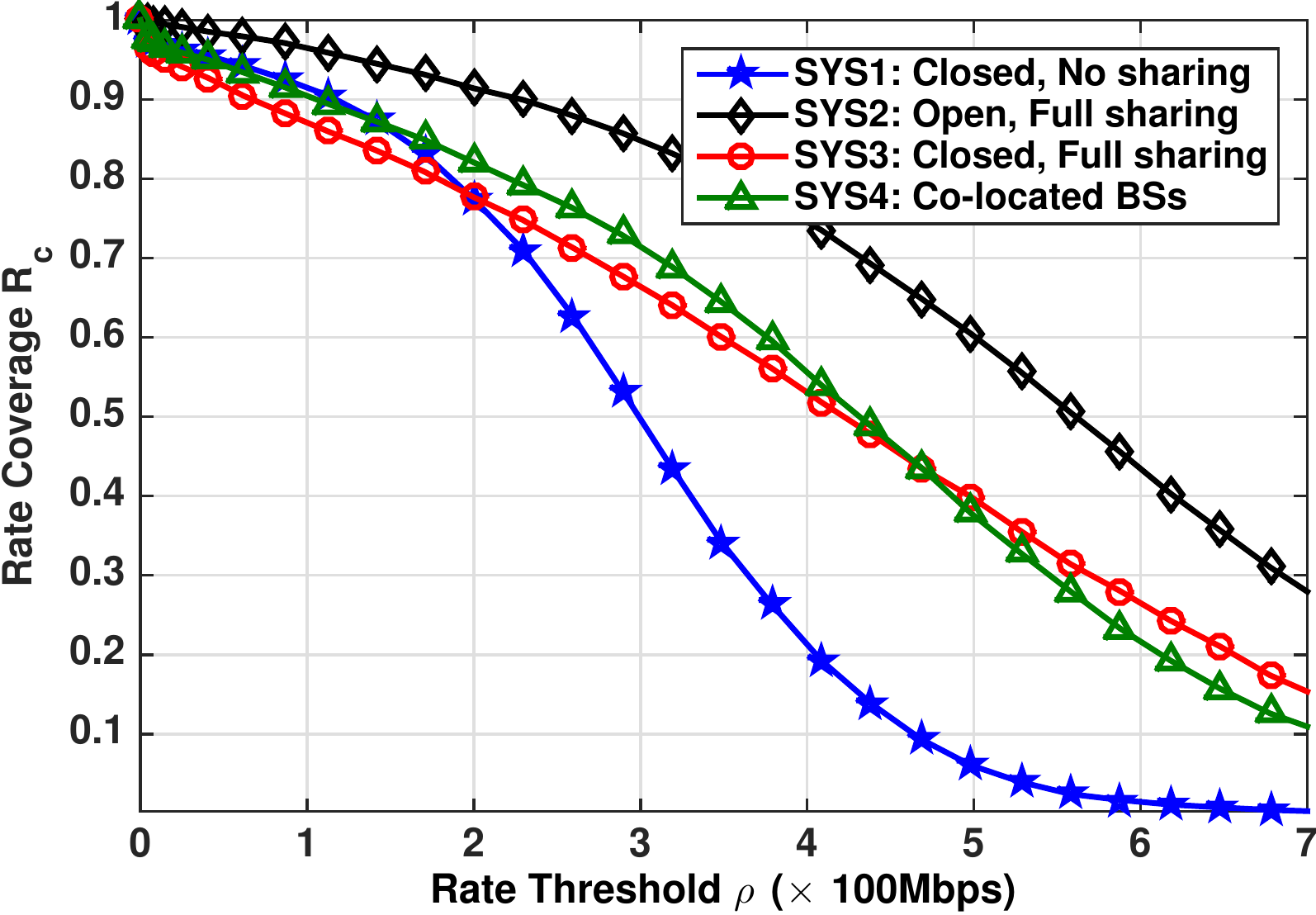}
\caption{Rate coverage in a two-network mmWave system  under partial loading with Rayleigh fading  for different cases.  Partial loading favors spectrum license sharing.}
\label{fig:rcploading}
\end{figure}

\begin{figure}[!ht]
\centering
\includefig{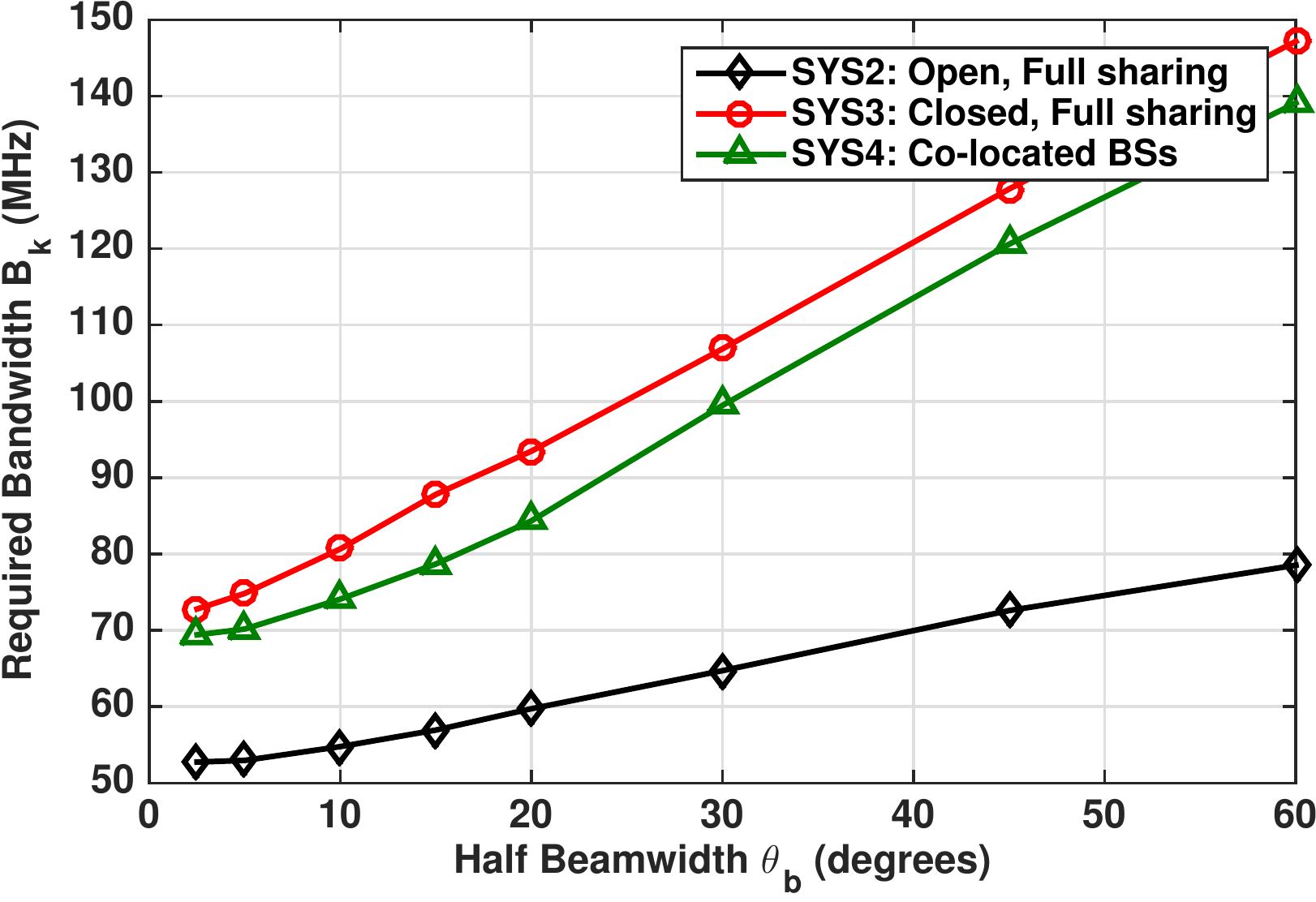}
\caption{Required bandwidth for each network  (with sharing of licenses) to achieve the same median rate achieved by the network with no sharing of spectrum with each network having 100MHz spectrum license. Sharing can reduce the license cost by more than 25\%}
\label{fig:reqbw}
\end{figure}

\textbf{Partial loading favors sharing:} \figurekeyword \ref{fig:rcploading}  compares the probability of rate coverage for four systems under partial loading with user density of $30/\text{km}^2$. It can be observed that due to reduced interference, System 3 (and 4) has even higher gain than System 1. \addvzerothree{In particular, System 3 has 40\% higher median rate than System 1 in partial loading case compared to  only 25\% gain in the previous case when user density was $200/\text{km}^2$.}

%
%

\textbf{Sharing  reduces  spectrum cost significantly:} Now, we compare the following two cases.  In the first case, each network owns a 100 MHz bandwidth exclusive license. This case is the same as System 1. In the second case, the networks share licenses completely and choose to buy just enough spectrum  to achieve the same median rate as in the first case. \figurekeyword \ref{fig:reqbw}  shows this required spectral bandwidth for each network. With a $10^\mathrm{o}$ beamwidth antenna, each network only needs to buy 75 MHz of bandwidth which would save 25\% of the license cost assuming linear pricing of the spectrum.

\begin{figure}[!t]
\centering
\iftoggle{SC}{
\includegraphics[width=0.570\textwidth, trim=0 0 0 0,clip=true]{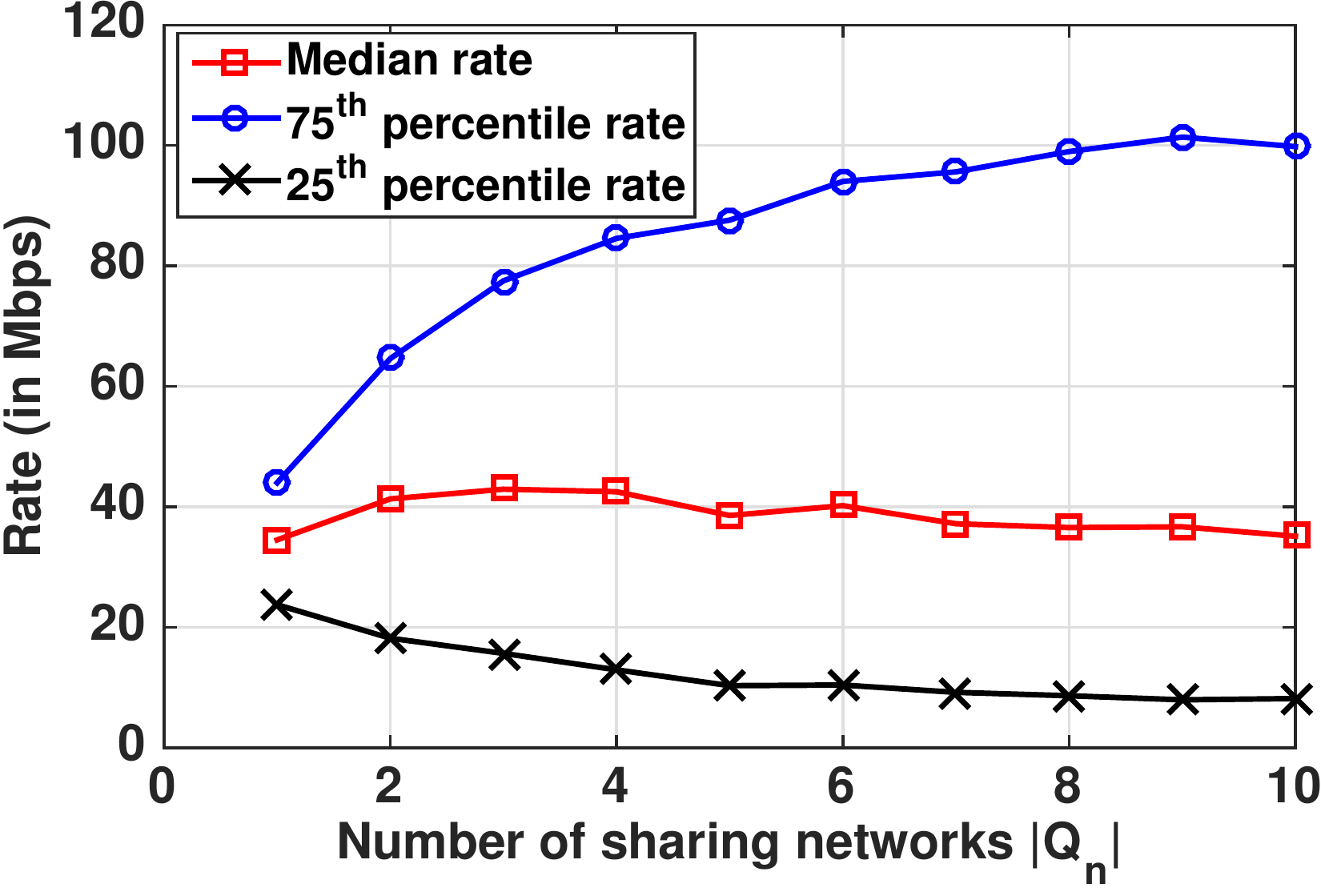}
}{
\includegraphics[width=.95\columnwidth, trim=0 0 0 0,clip=true]{figures/fig9}}
\vspace{-.15in}
\caption{Rate versus number of sharing networks in a mmWave cellular system with 10 networks. A trade-off between increasing the available bandwidth and  increasing interference is observed.}
\label{fig:numbervsrate}
\end{figure}

\addvzerofour{
\textbf{Optimal cardinality of sharing groups depends on the target rate:} Now, we consider a system with 10 operators with 50MHz bandwidth each and closed access. \figurekeyword  \ref{fig:numbervsrate} shows variation of the per-user rate for different percentiles  with respect to cardinality $|Q_n|$ of sharing group which is equal to the number of operators sharing licenses with network $n$. We can see that the 75\tho   percentile  rate increases with $|Q_n|$ while the 25\tho   percentile rate decreases. For the median rate, we see an increase  up to $|Q_n|$ = 3 and  then the median rate decreases. This trade-off is  due to the fact that as more  operators share their licenses, the total available bandwidth and the sum interference both increase. It can be observed that depending on the target performance, the optimal number of networks that should share their licenses varies.}

\begin{figure}[!ht]
\centering
\includefig{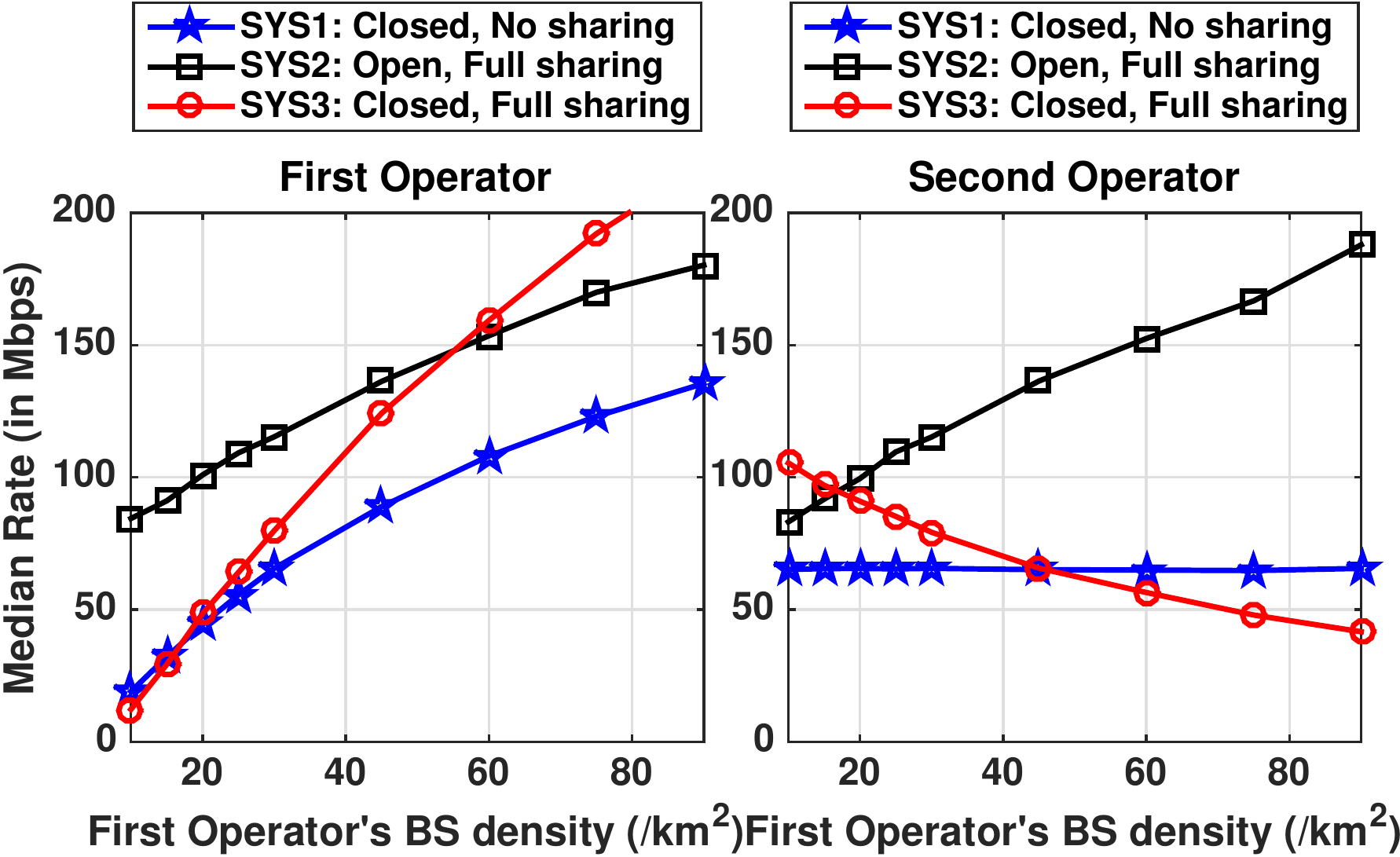} 
\caption{Variation of median rate with BS density of first operator in a two operator system. The BS density of second operator is fixed at 30/km$^{2}$. The user densities of both  operators are same and fixed. Both operators gain from  license sharing only when first operator density between 15 to 45/km$^{2}$. Spectrum license sharing is more beneficial for the operator with higher density.}
\label{fig:lambdachange}
\end{figure}

\begin{figure}[!ht]
\centering
\includefig{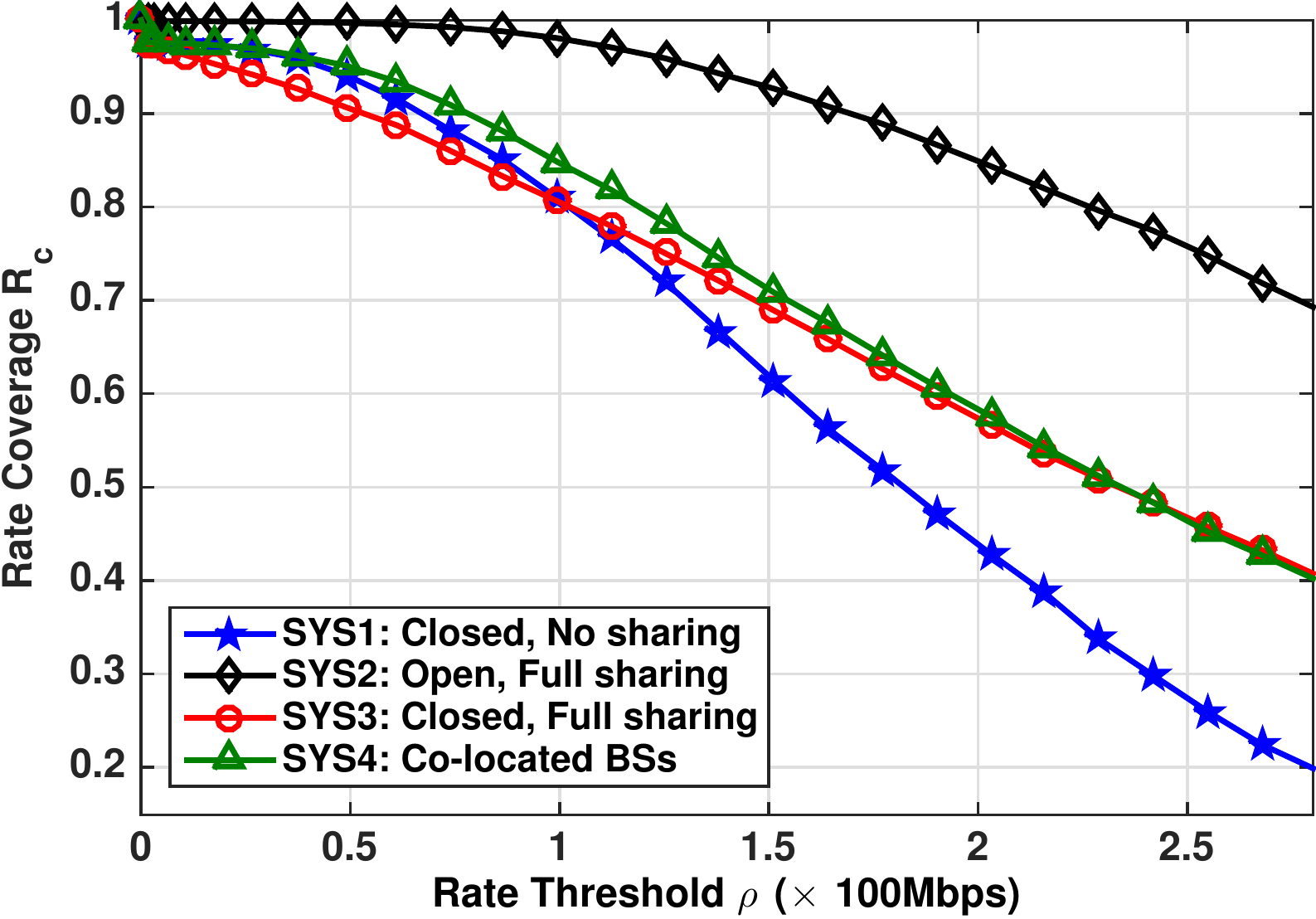} 
\caption{Rate coverage in a two-network mmWave system at 73 GHz frequency with Rayleigh fading  for different cases. Similar trends for spectrum sharing are observed for the 28 GHz and 73 GHz bands. }
\label{fig:rc73ghz}
\end{figure}

\textbf{Impact of asymmetry among operators:} We now consider the case when both operators are not identical. \figurekeyword \ref{fig:lambdachange}  shows the variation of median rate of the first and the second operators with respect to BS density of the first operator in a two-operator system with fixed user density of 200/km$^{2}$. The BS density of first operator is fixed at 30/km$^{2}$.  The gain from  license sharing  increases as the BS density increases for the operator with higher density and decreases for the other operator. We can observe that both operators can simultaneously gain from  license sharing only when both operators have similar BS densities which is between 15 to 45/km$^{2}$ here.

\textbf{Results for 73GHz band:} We also consider mmWave communication at 73GHz with 1 GHz bandwidth. The near-field path-loss gains are decreased by a factor of $10\log\mathrm{10}{(73/28)^2}=8.32\dB$ when compared with 28GHz for both LOS and NLOS.
 \figurekeyword \ref{fig:rc73ghz}  compares the rate coverage for four systems for 73GHz and shows similar trends as 28 GHz.  {Due to reduced interference, license sharing achieves slightly higher gain (28\%) compared to 28 GHz case ({\em i.e.} 25\%). }
 
\section{Conclusions}\label{Sec:Conclusions}
We have modeled a two-level architecture of a mmWave multi-operator system and derived the SINR and per-user rate distribution. We show that license sharing among operators improves system performance by increasing per-user rate. We conclude that it is economical for operators to share their spectrum licenses without increasing any overhead. We show that narrow beams play a key role in determining the feasibility of spectrum sharing. Since an increasing number of networks increases both the sum interference and bandwidth, the optimal cardinality of the sharing group will depend on the target rate.

This work would seem to have numerous extensions. First, it is worth investigating how multi-antenna techniques such as multiplexing including hybrid beamforming affect the insights about license sharing. Second, it is interesting to understand how other essential infrastructure including backhaul can be shared among networks reducing the cost further, especially in the case of co-located BSs.

\appendices
\section{Proof of Lemma \ref{lemma:laplace}} \label{lemmaproof:laplace}
The  interference from all LOS BSs of network $m$ at \UE~is given as 
\begin{align*}
I_{m\L}=\sum_{\x_{mj}\in \Phi'_{m,\L}
}
            {h_{mj}}{\|x_{mj}\|^{-\a_\L}}P_mC_\L
            G(\theta_{mj})
\end{align*}
where $\Phi'_{m,\L}=\Phi_{m,\L}\cap \bar{\Ball}\left(0,D^{ks}_{m\L}(x)\right)$ and $\bar{\Ball}\left(0,r\right)$ denotes the compliment of a ball of radius $r$ located at origin. This  is due to the fact that all interfering BSs are located outside the radius $D^{ks}_{m\L}(x)$. The Laplace transform of $I_{m\L}$ is given as
\begin{align*}
 \iftoggle{SC}{}{&}
 \laplace{I_{m\L}}(t)
 \iftoggle{SC}{}{ \\ }
 &= \expect{\exp{
        \left(-t\sum_{\x_{mj}\in \Phi'_{m\L}
        }
            {h_{mj}}{\|x_{mj}\|^{-\a_\L}}P_mC_\L
            G(\theta_{mj})
        \right)
        }
    }.
\end{align*}
Now, using the PGFL of PPP \cite{Stoyanbook}, the Laplace transform can be written as
\begin{align*}
\laplace{I_{m\L}}(t)  =& 
\exp{}
        \left(
            -2\pi\lambda_{m}
            \int_{D^{ks}_{m\L}(x)}^{\infty}
            p(y)
            \iftoggle{SC}{}{\right.\\&\ \ \left.}
            \left(
            1-\expect{
                e^{-t {h G(\theta)P_mC_\L}
                {y^{-\a_\L}}}}
            \right)
            y\intd y
        \right)
        {}.
        \end{align*}
 Now, using the moment generating function (MGF) of exponentially distributed $h$ and pdf of uniformly distributed $\theta$, we get
\begin{align*}
  \laplace{I_{m\L}}(t)      =& 
  \exp{}
        \left(
            -\lambda_{m}
            \int_{D^{ks}_{m\L}}^{\infty}
            p(y)
            \iftoggle{SC}{}{\right.\\&\ \ \left.}
           \left(2\pi-\int_0^{2\pi}{
                \frac{d\theta}
                {1+t
                {G(\theta)P_mC_\L}{y^{-\a_\L}}
                }}      
            \right)
            \iftoggle{SC}{}{\vphantom{ \int_{D^{ks}_{m\L}}^{\infty}}}
            y\intd y
        \right)
        {}.
\end{align*}
  Now, integrating with $\theta$ and then using the definition of function $F_\L(\cdot)$, we get
  $\iftoggle{SC}{}{\laplace{I_{m\L}}(t)=}$
 \begin{align*}
 \iftoggle{SC}{\laplace{I_{m\L}}(t)=}{}
&\exp{
        \left(
            -\lambda_{m}
            {\int_{D^{ks}_{m\L}}^{\infty}\int_{0}^{2\pi}
            {
                \frac{t
                {G(\theta)P_mC_\L}{y^{-\a_\L}}}
                     {1+t
            {G(\theta)P_mC_\L}{y^{-\a_\L}}}
            }\intd\theta
            e^{-\beta y}y\intd y }
        \right)
        }\\
&=\exp{}
        \left(
            -2\lambda_{m}
            \int_{D^{ks}_{m\L}}^{\infty}\left(
            {\thetab
                \frac{t
                {G_1P_mC_\L}{y^{-\a_\L}}}
                     {1+t
            {G_1P_mC_\L}{y^{-\a_\L}}}
            }
            \iftoggle{SC}{}{\right.\right.\\&\hspace{0.5in}\ \ \ \left.\left.}
            +
             {
            (\pi-\thetab)
                \frac{t
                {G_2P_mC_\L}{y^{-\a_\L}}}
                     {1+t
            {G_2P_mC_\L}{y^{-\a_\L}}}
            }\right)
            e^{-\beta y}y\intd y
        \right)
        {}\\
&=\exp{}
        \left(
            -2\lambda_{m}\left[
            \thetab
            F_\L\left(\beta,\a_\L,tG_1P_mC_\L,D^{ks}_{m\L}(x)\right)
          \iftoggle{SC}{}{  \right.\right.\\ & \equalspace\equalspace \left.\left.}
            +(\pi-\thetab)
            F_\L\left(\beta,\a_\L,tG_2P_mC_\L,D^{ks}_{m\L}(x)\right)
            \right]
        \right).
 \end{align*}
The Laplace transform of the interference from the NLOS BSs can be computed similarly. 
 
 \section{Proof of Lemma \ref{lemma:laplacecolocs} }
 \label{lemmaproof:laplacecolocs}
 The Laplace transform of interference in \eqref{eq:IntExpcolocs} can be computed as $\laplace{I}(t)=$
 \begin{align*}
&\expects{h,\theta}{e^{-tx^{-\alpha_s}C_s\sum_{m\in Q_k\setminus\{k\}}P_m h_{mi} G(\theta_{mi})}}
\iftoggle{SC}{}{\\&}
\prod_{p=\L,\N}\expects{\Phi_p,h,\theta}{e^{-t\sum_{\x_j\in \Phi_p\setminus\{\x_i\}}x_j^{-\alpha_p}C_p\sum_{m\in Q_k}P_mh_{mj}G(\theta_{mj})}}.
\end{align*}
Now, using the PGFL of PPP and independence of $h_{mi}$'s and $\theta_{mi}$'s,  $\laplace{I}(t)$ can be written as
\begin{align*}
\laplace{I}(t) =&
\prod_{{m\in Q_k\setminus\{k\}}}\expects{h,\theta}{e^{-tx^{-\alpha_s}C_sP_m h_{mi} G(\theta_{mi})}}
\iftoggle{SC}{\\&}{\\&}
\prod_{p=\L,\N}\exp{}\left(-2\pi\lambda\int_{D^s_p(x)}^\infty p(y)
\iftoggle{SC}{}{\right.\\&\left.}
\expects{h,\theta}{1-e^{-ty^{-\alpha_p}C_p\sum_{m\in Q_k}P_mh_{m}G(\theta_{m})}}y\intd y\right){}.
\end{align*}
Now, using the MGF of exponentially distributed  $h_{mi}$'s in first product term and the independence of $h_m$'s in the second product term, we get
\begin{align*}
\laplace{I}(t) =&\prod_{{m\in Q_k\setminus\{k\}}}\expects{\theta}{\frac1{1+tx^{-\alpha_s}C_sP_m  G(\theta_{mi})}}
\iftoggle{SC}{\\&}{\\=&}
\prod_{p=\L,\N}\exp{}\left(-2\pi\lambda\int_{D^s_p(x)}^\infty p(y)
\iftoggle{SC}{}{\right.\\&\left.}
\expects{\theta}{1-\prod_{m\in Q_k}\expects{h}{ e^{-ty^{-\alpha_p}C_pP_mh_{m}G(\theta_{m})}}}y\intd y\right){}.
\end{align*}
Now, using the MGF of exponentially distributed  $h_{m}$'s,  we can write\begin{align*}
\expects{h}{ e^{-ty^{\alpha_p}C_pP_mh_{m}G(\theta_{m})}}&=
1/\left({1+ty^{-\alpha_p}C_pP_mG(\theta_{m})}\right).
\end{align*}
Since $G(\theta_m)$'s are discrete random variables with $\prob{G(\theta_m)=G_1}=\thetab/\pi$ and  $\prob{G(\theta_m)=G_2}=1-\thetab/\pi$, $\laplace{I}(t)$ can be further written as
\begin{align*}
&\laplace{I}(t) 
=\prod_{{m\in Q_k\setminus\{k\}}}\left(\frac{\thetab/\pi}{1+tx^{-\alpha_s}C_sP_m G_1}+\frac{(\pi-\thetab)/\pi}{1+tx^{-\alpha_s}C_sP_m G_2}\right)\\
&\ \times \prod_{p=\L,\N}\exp{}\left(-2\pi\lambda\int_{D^s_p(x)}^\infty p(y)
\iftoggle{SC}{}{\right.\\&\left.}
\left({1-\prod_{m\in Q_k}
{\left(\frac{\thetab/\pi}{1+ty^{-\alpha_p}C_pP_mG_1}+\frac{(\pi-\thetab)/\pi}{1+ty^{-\alpha_p}C_pP_mG_2}\right)}
}\right)y\intd y\right){}
\end{align*}
which proves the Lemma.

\section{Proof of the Laplace Transform for Co-located BS under Partial Loading}\label{appen:3}
The Laplace transform of interference in partial loading case  \eqref{eq:IntExpcolocsPL} is given as $\laplace{I}(t)=$
 \begin{align*}
&\expects{h,\theta}{e^{-tx^{-\alpha_s}C_s\sum_{m\in Q_k\setminus\{k\}}P_m h_{mi} G(\theta_{mi})\delta_{mi}}}
\iftoggle{SC}{}{\\&}
\prod_{p=\L,\N}\expects{\Phi_p,h,\theta}{e^{-t\sum_{\x_j\in \Phi_p\setminus\{\x_i\}}x_j^{-\alpha_p}C_p\sum_{m\in Q_k}P_mh_{mj}G(\theta_{mj})\delta_{mi}}}.
\end{align*}
Since $\delta_{mi}$'s are Bernoulli random variables with $\prob{\delta_{mi}=0}=\kappa(0)$,  $\laplace{I}(t)$ can be  written as
\begin{align*}
\laplace{I}(t)=&
\prod_{{m\in Q_k\setminus\{k\}}}\expects{h,\theta}{\kappa(0)+(1-\kappa(0))e^{-tx^{-\alpha_s}C_sP_m h_{mi} G(\theta_{mi})}}
\iftoggle{SC}{\\&}{\\&}
\prod_{p=\L,\N}\exp{}\left(-2\pi\lambda\int_{D^s_p(x)}^\infty p(y)
\iftoggle{SC}{}{\right.\\&\hspace{0.2in}\left.}
\expects{h,\theta,\delta}{1-e^{-ty^{-\alpha_p}C_p\sum_{m\in Q_k}P_mh_{m}G(\theta_{m})\delta_m}}y\intd y\right){}\\
=&\prod_{{m\in Q_k\setminus\{k\}}}\expects{\theta}{\kappa(0)+(1-\kappa(0))\frac1{1+tx^{-\alpha_s}C_sP_m  G(\theta_{mi})}}
\iftoggle{SC}{\\&}{\\&}
\prod_{p=\L,\N}\exp{}\left(-2\pi\lambda\int_{D^s_p(x)}^\infty p(y)\expects{\theta}{1-\prod_{m\in Q_k}
\left(\kappa(0)
\iftoggle{SC}{}{\right.\right.\right.\\&\hspace{0.2in}\left.\left.\left.}
+(1-\kappa(0))
\expects{h}{ e^{-ty^{\alpha_p}C_pP_mh_{m}G(\theta_{m})}}
\right)
\iftoggle{SC}{}{\vphantom{\prod_{m\in Q_k}}}
}y\intd y 
\iftoggle{SC}{}{\vphantom{\int_{D^s_p(x)}^\infty}}
\right){}
\end{align*}
which can be  simplified further following the same steps as Appendix \ref{lemmaproof:laplacecolocs} to get \eqref{eq:laplacecolocsPL}.

\end{document}